\documentclass[prd,reprint, nofootinbib,amsmath,amssymb,amsfonts,aps,showpacs,
superscriptaddress,eqsecnum, floatfix, longbibliography]{revtex4-1}
\usepackage[colorlinks=true,allcolors=blue]{hyperref}
\usepackage{graphicx}

\begin{document}

\title{Spin memory effect for compact binaries in the post-Newtonian 
approximation}
\author{David A.~Nichols}
\email{d.nichols@astro.ru.nl}
\affiliation{Department of Astrophysics, Faculty of Science, Radboud
University Nijmegen, P.O.~Box 9010, 6500 GL Nijmegen, The Netherlands}

\begin{abstract}
The spin memory effect is a recently predicted relativistic phenomenon in
asymptotically flat spacetimes that become nonradiative infinitely far in the
past and future.
Between these early and late times, the magnetic-parity part of the time
integral of the gravitational-wave strain can undergo a nonzero change;
this difference is the spin memory effect.
Families of freely falling observers around an isolated source
can measure this effect, in principle, and fluxes of angular momentum per unit
solid angle (or changes in superspin charges) generate the effect.
The spin memory effect had not been computed explicitly for astrophysical 
sources of gravitational waves, such as compact binaries.
In this paper, we compute the spin memory in terms of a set of radiative
multipole moments of the gravitational-wave strain.
The result of this calculation allows us to establish the following results
about the spin memory:
(i) We find that the accumulation of the spin memory behaves in a qualitatively
different way from that of the displacement memory effect for nonspinning, 
quasicircular compact binaries in the post-Newtonian approximation: the spin 
memory undergoes a large secular growth over the duration of the inspiral, 
whereas for the displacement effect this increase is small.
(ii) The rate at which the spin memory grows is equivalent to a 
nonlinear, but nonoscillatory and nonhereditary effect in the gravitational 
waveform that had been previously calculated for nonspinning, quasicircular 
compact binaries.
(iii) This rate of build-up of the spin memory could potentially be detected 
by future gravitational-wave detectors by carefully combining the measured 
waveforms from hundreds of gravitational-wave detections of compact binaries.
\end{abstract}

\maketitle

\section{\label{sec:intro} Introduction}

\subsection{Overview of the displacement memory effect}

The displacement memory effect is a well studied property of asymptotically 
flat spacetimes that limit to nonradiative solutions infinitely far in the 
past and future.
The effect is characterized by the gravitational-wave strain undergoing a 
change between these early and late times.
Nearby freely falling observers can measure this effect through their
geodesic deviation.
In principle, any isolated source that radiates (effective) energy density
in massless (or gravitational) fields asymmetrically, or that has a change 
in its supermomentum charges, can produce the effect.
It is of great interest to know whether astrophysical sources that exist in 
our observable universe can generate the effect with a sufficient amplitude 
so that it might be detected by current or upcoming gravitational-wave 
experiments.

Zel'dovich and Polnarev~\cite{Zeldovich1974} made an important first step in 
this direction when they computed, in linearized gravity, the 
displacement memory arising from the scattering of stars.
Related calculations of the displacement memory from gravitational 
bremsstrahlung were performed soon 
after~\cite{Turner1977,Turner1978a,Kovacs1978}.
Turner also noted that neutrino emission from supernovae would be a source of 
the displacement memory effect~\cite{Turner1978b}.
These early calculations of the effect were all in the context of linear
gravity.
Christodoulou~\cite{Christodoulou1991} found that the effective energy per 
solid angle radiated in gravitational waves could also be a source of the
displacement memory, in principle.
Wiseman and Will~\cite{Wiseman1991} calculated this nonlinear displacement 
memory for compact binaries, and Blanchet and Damour~\cite{Blanchet1992} 
computed the effect in the post-Newtonian-expanded, multipolar post-Minkowski 
approximation (henceforth, the PN approximation, for short).
They were able to confirm that compact binaries do have a nontrivial memory
effect.
The displacement memory effect has subsequently been computed to a high order 
in the PN approximation \cite{Favata:2008yd} and in numerical relativity 
simulations \cite{Pollney2010} (comparisons between numerical and analytical
calculations of the displacement memory are given in~\cite{Cao:2016maz}).

With a range of astrophysical sources that generate the displacement memory and
with the improvement in the sensitivity of current gravitational-wave 
detectors, observing the displacement memory is now an active goal of 
gravitational-wave research.
Initial strategies to detect the displacement memory signal were proposed as 
long as roughly thirty years ago \cite{Braginsky1985,Braginsky1987}.
Pulsar timing experiments have performed searches for the displacement memory,
though they have not yet found evidence for the 
phenomenon~\cite{Wang2015,Arzoumanian2015}.
In addition, it will be possible to use LIGO~\cite{Abbott:2007kv} (alone or in
concert with Virgo~\cite{TheVirgo:2014hva} and KAGRA~\cite{Aso:2013eba}) to 
detect the accumulation of the displacement memory by coherently combining on 
the order of one-hundred memory signals from gravitational-wave detections of 
binary black holes~\cite{Lasky:2016knh}.
Future ground-based gravitational-wave detectors will be able to observe
the displacement memory in even greater detail, and space-based 
interferometers, such as the LISA mission~\cite{Audley:2017drz}, would be 
capable of detecting the memory signal from individual 
supermassive-binary-black-hole mergers~\cite{Favata:2009ii}.

The displacement memory also has an interesting connection to 
the symmetry group of asymptotically flat spacetimes, the Bondi-Metzner-Sachs 
group~\cite{Bondi1962,Sachs1962a,Sachs1962b}.
This group has a similar structure to the Poincar\'e group, in the
sense that it is the semidirect product of an abelian group with a
group isomorphic to the proper, orthochronous Lorentz group; however, 
rather than having a four-dimensional group of translations, it has an
infinite-dimensional group of supertranslations (which include the 
four-dimensional group of ordinary translations).
The additional supertranslations, roughly, can be thought of as 
``angle-dependent'' translations.
The displacement memory is related to the supertranslation needed to 
reach a particular frame at late times from a related frame at early times 
(see, e.g.,~\cite{Strominger:2014pwa,Flanagan:2015pxa} for recent 
descriptions of this property).
Alternatively, it can be thought of as the transformation needed to transform
a ``preferred'' Poincar\'e group at early times to a different preferred
Poincar\'e group at late times (e.g.,~\cite{Ashtekar:2014zsa}).
Finally, there is one additional interesting relationship between the memory
and asymptotic symmetries: the changes in charges conjugate to the 
supertranslations, the supermomentum charges, are also a source of the 
displacement memory.

\subsection{Summary of the spin memory}

The history of the spin memory has unfolded following a somewhat different 
path.
Its discovery was inspired (at least in part) by the work of Barnich and 
Troessaert~\cite{Barnich2009,Barnich2010} (see also~\cite{Banks2003}).
In these papers, they proposed that the symmetry algebra of asymptotically
flat spacetimes be enlarged to incorporate all symmetry vector fields, 
including those that have analytic singularities.
This extends the Lorentz part of the algebra to the Virasoro algebra with
vanishing central charge.
The additional generators were dubbed ``super-rotations,'' because they
can be thought of as generalizations of the Lorentz vector fields (very 
roughly, they are a certain type of angle-dependent rotations and boosts).

Several years later, Pasterski \textit{et al}.~\cite{Pasterski:2015tva} 
observed that there is a new memory effect that is related to the 
super-rotation symmetry, though in a somewhat different way from how the 
displacement memory and the supertranslation symmetry are related 
(see also~\cite{Flanagan:2015pxa}).
This new memory was dubbed the ``spin memory,'' because it is sourced by 
asymmetric changes in the (effective) angular momentum per unit solid angle 
radiated to null infinity in massless (and gravitational) fields, as well 
as changes in the superspin charges (the magnetic-parity part of the charges 
conjugate to the super-rotation vector fields).
However, it has no relation to a spacetime super-rotated from a specific
early-time frame, because such spacetimes are not asymptotically flat in the
traditional sense (see, e.g.,~\cite{Strominger:2016wns})

There are (at least) two idealized observational methods by which the spin
memory can be measured: (i) Construct a set of Bondi coordinates 
around an isolated source and require that an observer follow an accelerating 
worldline of constant spatial Bondi coordinates. 
Have this observer carry a Sagnac detector and have her measure the 
time-integrated change in the Sagnac effect between early and late times.
The resulting change is the spin memory~\cite{Pasterski:2015tva}.
(ii) Consider a family of freely falling observers around an isolated source.
Have them measure their geodesic deviation from their neighboring observers 
and allow them to compare their results with these other observers around the 
source.
From these measurements they can extract the magnetic-parity portion of the 
sky pattern of the time-integrated strain that defines the spin 
memory~\cite{FlanaganInPrep}.
Both (i) and (ii) are in fact nonlocal measurements, because some nonlocal
information is hidden in the procedure to define Bondi coordinates.

There is, therefore, compelling theoretical evidence that, in principle, the
spin memory exists and is measurable.
However, there are neither any calculations of the magnitude of the effect
from astrophysical sources, nor are there explicit strategies to detect the
spin memory using current or future gravitational-wave detectors.
This paper is a first step towards addressing these gaps in our understanding
of the spin memory.
Our focus will be on compact binary systems in quasicircular orbits which,
given the recent detections of black-hole binaries by the LIGO-Virgo 
Collaboration~\cite{TheLIGOScientific:2016pea}, are a particularly relevant 
and promising source.

Computing the spin memory in full generality from such sources would be 
challenging and would likely require the computational infrastructure of 
Cauchy numerical relativity simulations combined with Cauchy-characteristic 
extraction (e.g.,~\cite{Bishop:1996gt,Reisswig:2009us,Handmer:2016mls}).
We will not attempt such a large undertaking in this paper.
Instead, we will derive an expression for the spin memory in terms of a set 
of radiative multipole moments of the gravitational-wave strain.
Within the PN approximation, we can then relate the radiative moments to a set
of source moments and thereby obtain quantitative values for the magnitude
of the growth of the spin memory from specific types of astrophysical sources 
(in particular, compact binaries in quasicircular orbits).

\subsection{Methodology for computing memory effects}

In the Bondi framework, the displacement memory can be computed from one 
component of Einstein's equations (which turns out to contain equivalent 
information to the flux law for the supermomentum charges).
The contributions to the displacement memory can be categorized into three 
types: changes in the supermomentum charges (the ordinary displacement 
memory in the nomenclature of~\cite{Bieri:2013ada}), fluxes of the 
stress-energy tensor of matter fields (the matter part of the null displacement
memory), and fluxes of the effective stress-energy of gravitational fields 
(the nonlinear null displacement memory).
These contributions are defined in terms of quantities far from an isolated
source without reference to quantities in the interior of the spacetime.
The displacement memory effect is then the limiting change of the
gravitational-wave strain produced by these three contributions, between 
the asymptotically nonradiative regions of the spacetime infinitely far in the 
past and future.

The PN approximation takes a different approach to defining the displacement
memory effect.
Because the PN approach perturbatively relates the dynamics of a source to its
emitted gravitational waves via an asymptotic matching calculation, it cannot 
describe the complete evolution of a compact binary; the errors in the 
perturbative expansion grow large when the binary's separation is small.
It is not possible to use the PN approximation until the binary settles into
the asymptotically nonradiative state in the future that is needed to define 
the displacement memory effect.
Instead a proxy for the displacement memory is defined in PN theory by 
a specific piece of the PN-expanded wave-zone metric perturbation at a fixed
time, which is not instantaneous (nonhereditary) and nonlinear and resembles 
the retarded integral of the effective energy density of the gravitational 
field (see, e.g.,~\cite{Blanchet:2006zz}).
Specifically, this contribution is the nonlinear part of the null displacement 
memory, in the terminology of~\cite{Bieri:2013ada}.
It is also this piece of the metric that serves as the displacement memory 
signal that can be detected by gravitational-wave detectors 
(such as LIGO~\cite{Lasky:2016knh}) rather than the 
formal asymptotic change in the gravitational-wave strain.

In the Bondi framework, the spin memory has contributions from three types
of sources that are analogous to those of the displacement memory: there is
an ordinary part from changes of the superspin charges, a matter part of
the null spin memory from the flux of angular momentum per solid angle in 
matter fields, and a nonlinear null spin memory from the flux of effective
angular momentum per solid angle in gravitational waves.
In the PN framework, the spin memory has not been computed, and there
is not yet an established definition of what the spin memory observable is
in this approximation.
We will define the spin memory in this context by its nonlinear null part
evaluated at a finite retarded time, rather than infinitely far in the 
future (analogously to how the displacement memory is defined in PN theory).
The features of this definition of the spin memory are described next.

\subsection{Summary of results}

Our results are based on an expression for the spherical-harmonic modes of the 
spin memory observable that we compute in this paper in terms of the radiative
multipole moments of the gravitational-wave strain.
From this expression, we can determine the spin memory for nonspinning 
quasicircular compact binaries in the PN approximation.
For these systems, we will denote the total mass of the binary by $M$ and its 
orbital frequency by $\omega$.
We will form the dimensionless PN quantity $x=(M\omega)^{2/3}$ (in geometric 
units in which $G=c=1$).
Our calculations show that the spin memory is contained predominantly in the
$l=3$, $m=0$ mode (for a coordinate frame in which the binary has its orbital 
angular momentum oriented along the $z$ direction).
This mode scales as $x^{-1/2}$; namely, it gradually accumulates over the 
long inspiral to have a quite significant effect.\footnote{As we discuss in 
greater detail in Sec.~\ref{subsec:SpinMemFreq}, this dependence of the spin
memory on frequency is in line with that of the radiated angular momentum 
at leading Newtonian order.
It will require that we consider binaries that are formed at a finite
distance in the past (so that their orbital frequency never becomes zero). 
All binaries in the observable universe fall into this class; in this
sense, it is neither a significant physical restriction nor an indication
of a problematic feature of the spin memory effect.}
One should contrast this with the displacement memory, which scales as $x$ 
(it grows most rapidly when the binary is most relativistic and has small 
contributions from the early stages of the inspiral).

While this difference in scaling at low orbital frequencies is indeed a 
curiosity, it is difficult to envision how to observe this feature directly.
The gradual accumulation of the spin memory takes place over such a long 
timescale that for any individual event, it is difficult to envision a way
to measure this effect.
LIGO, for example, is sensitive to just the late stages of the inspiral (and 
the merger and ringdown) of stellar-mass compact binaries, and it measures
the strain, not its time integral.
This would seem to pose a challenge for detecting the spin memory.
However, we are able to show that the rate at which the spin memory 
accumulates is equivalent to a nonlinear, nonhereditary, and nonoscillatory
component of the gravitational waveform that had previously been discovered
at 2.5 PN order for nonspinning compact binaries~\cite{Arun:2004ff}.
For this reason, we will subsequently call this term the ``spin memory mode.''
This quantity is conceivably detectable, although its high PN order would 
make it challenging to do so.

To estimate the spin memory mode's detectability, we use improvised (though 
nonrigorous) waveforms created by inserting numerical-relativity 
gravitational-waveform modes from the SXS waveform catalog~\cite{SXS:catalog}
into the leading PN-order multipolar expression we derive for the spin memory.
For individual binary mergers, we find that the signal-to-noise ratio (SNR) 
in the spin memory component of the waveform is below the detection threshold.
However, by coherently combining the spin memory signals using a procedure
similar to that described in~\cite{Lasky:2016knh}, we suspect that the effect
could be measurable by future ground-based detectors like the Einstein 
Telescope (see, e.g.,~\cite{Punturo:2010zz}).

\subsection{Outline of this paper}

We structure the remainder of our paper in the following way:
In Sec.~\ref{sec:review}, we describe the basic elements of the PN 
approximation and the Bondi framework.
It is here that we describe a way to incorporate a set of PN radiative moments
into a Bondi frame.
As a warm up, we show in Sec.~\ref{sec:disp_memory} that we can use this 
PN-expanded Bondi frame to correctly predict the multipolar expression for 
the PN computation of the displacement memory, including the correct 
leading-order expression for the PN displacement memory of compact binary 
systems.
With this result completed, we then turn to the calculation of the nonlinear
part of the null spin memory in Sec.~\ref{sec:spin_memory}.
We give both a general expression in terms of the radiative moments of the
gravitational-wave strain and a leading-order result for compact binaries in 
this section.
We also show how the rate of accumulation of the spin memory corresponds to
the 2.5PN nonlinear, nonoscillatory, and nonhereditary part of the 
gravitational waveform (the spin memory mode).
In Sec.~\ref{sec:detection}, we estimate the SNR of the spin memory mode
in current and future ground-based detectors.
We sketch a procedure for coherently adding spin memory signals to allow
for the detection of this phenomenon.
In Sec.~\ref{sec:conclude}, we conclude.
Appendix~\ref{sec:harmonic_relations} contains certain properties of 
the vector and tensor spherical harmonics that appear in our calculations.
As a check of our calculation of the spin memory, we show in 
Appendix~\ref{sec:AngMomFlux} that we recover the same expression for the
flux of angular momentum flux expanded in multipole moments of the 
gravitational-wave strain that was derived in~\cite{Thorne1980} using very
different methods.
Finally, in Appendix~\ref{sec:SpinMemNR}, we compare the spin memory mode
in a numerical simulation with our improvised, approximate waveform model.

Throughout this paper, we use units in which $G=c=1$, although we will on
occasion denote relative PN errors with expressions of the form $O(c^{-n})$
where $c$ is the speed of light and $O(G^m)$ for Newton's constant $G$.
Our conventions for the metric and curvature tensors follow those given 
in Wald~\cite{Wald:1984rg}.

\section{\label{sec:review} Elements of the Bondi and PN frameworks}

In the two parts of this section, we introduce the minimal elements of the
PN approximation and the Bondi framework that we will need for this paper.
More detailed reviews of PN theory can be found, for example, 
in~\cite{Blanchet:2006zz}.
The textbook by Stewart~\cite{Stewart:1990uf}, for example, has significantly 
more information about asymptotically flat spacetimes, as does the recent
review article~\cite{Madler:2016xju}.

\subsection{\label{subsec:PN} PN approximation}

Our focus will be primarily on the asymptotic gravitational waveform within
the PN approximation, and our emphasis in this section, therefore, will 
largely fall on this aspect of PN theory.

\subsubsection{General aspects of PN theory}

The gravitational waveform in the PN approximation is encoded within
a symmetric, trace-free and transverse tensor $h_{ij}^{\mathrm{TT}}$.
We are using the notation that the Latin indices $i$ and $j$ refer to spatial
indices in inertial Minkowski coordinates.
The transverse-traceless metric perturbation is a function of two sets of 
radiative multipole moments, which are often called the mass and current 
moments.
We will expand $h_{ij}^{\mathrm{TT}}$ in sets of second-rank, ``pure-spin'' 
tensor harmonics of electric and magnetic parity.
The definitions of these harmonics and their relationship to the spin-weighted
spherical harmonics are given in Appendix~\ref{sec:harmonic_relations} 
(related definitions and relations of pure-spin vector and third-rank 
tensor harmonics are also given therein).

The waveform can be written as
\begin{equation}
h_{ij}^{\mathrm{TT}} = \frac 1 r \sum_{m,l\geq 2} [U_{lm}(u) T^{(e),lm}_{ij} + 
V_{lm}(u) T^{(b),lm}_{ij}] + O(1/r^2)\, ,
\label{eq:hijTT}
\end{equation}
where $U_{lm}(u)$ are the radiative mass moments and 
$V_{lm}(u)$ are the radiative current moments.
The coordinate $u=t-r$ is the retarded time.
Expressions for the tensor harmonics $T^{(e),lm}_{ij}$ and $T^{(b),lm}_{ij}$ 
can be obtained from Eq.~\eqref{eq:TABharmonics} after applying an 
appropriate transformation to Cartesian coordinates from spherical polar
coordinates.
The function $h_{ij}^{\mathrm{TT}}$ is real, whereas the second-rank, 
pure-spin tensor harmonics and the radiative mass and current moments are 
all complex quantities.
Because the tensor harmonics satisfy the complex-conjugate relationships
\begin{equation}
\bar T^{(e),lm}_{ij} = (-1)^m T^{(e),l-m}_{ij} \, , \qquad 
\bar T^{(b),lm}_{ij} = (-1)^m T^{(b),l-m}_{ij} \, ,
\end{equation}
the radiative mass and current moments must have a similar relation
\begin{equation}
\bar U_{lm} = (-1)^m U_{l-m} \, , \qquad 
\bar V_{lm} = (-1)^m V_{l-m} \, ,
\end{equation}
to ensure that $h_{ij}^{\mathrm{TT}}$ is real.

The moments $U_{lm}(u)$ and $V_{lm}(u)$ can also be related to the 
spin-weighted spherical-harmonic expansion of the complex waveform 
\begin{equation}
h = h_+ - i h_\times \, , 
\end{equation}
where $h_+$ and $h_\times$ are the real-valued gravitational-wave 
polarizations.
They are defined by projecting the strain $h_{ij}^{\mathrm{TT}}$ onto 
vectors in an orthonormal spatial dyad in the direction orthogonal to the 
propagation of the gravitational waves (see the end of 
Appendix~\ref{sec:harmonic_relations} for details).
The relevant relationship is
\begin{equation}
h =  \sum_{l,m} h_{lm}(u) {}_{-2}Y_{lm} \, ,
\end{equation}
where
\begin{equation}
h_{lm} = \frac 1{r\sqrt 2}(U_{lm} - i V_{lm}) \, .
\end{equation}
This expression can also be inverted to solve for the radiative mass and
current moments in terms of the modes $h_{lm}$.
The result is
\begin{subequations}
\begin{align}
U_{lm} = & \frac{r}{\sqrt 2} [h_{lm} +(-1)^{-m} \bar h_{l-m}] \, , \\
V_{lm} = & i \frac{r}{\sqrt 2} [h_{lm} - (-1)^{-m} \bar h_{l-m}] \, ,
\end{align}
\label{eq:UVfromhlm}%
\end{subequations}
which will be useful for relating analytical- and numerical-relativity
quantities in Appendix~\ref{sec:SpinMemNR}.

Finally, the radiative moments can be related to a set of source multipole
moments (in practice, this is done through an intermediate set of
canonical moments; see, e.g.,~\cite{Blanchet:2006zz} for details).
At leading order in the PN expansion, the radiative multipoles are equal
to $l$ derivatives of the source multipoles $I_{lm}$ and $J_{lm}$:
\begin{equation}
U_{lm} = I_{lm}^{(l)} + O(G) \, , \qquad
V_{lm} = J_{lm}^{(l)} + O(G) \, .
\label{eq:RadiativeToSource}
\end{equation}
We use $O(G)$ to mean there is a relative (not an absolute) error of order $G$.
We will employ the relations~\eqref{eq:RadiativeToSource} in the next part 
to express the radiative multipoles in terms of the total mass, symmetric mass 
ratio, and orbital frequency of a compact binary.
With the expression for the spin memory in terms of the radiative multipole 
moments given in Sec.~\ref{sec:spin_memory} and the 
expressions~\eqref{eq:RadiativeToSource}, one can compute the spin memory for
any source of one's choosing at leading PN order.
We will focus on nonspinning compact binary sources in quasicircular orbits
in this paper.

\subsubsection{PN theory for compact binaries in quasicircular orbits}

Consider a nonspinning, quasicircular compact binary in a coordinate
system with the orbital angular momentum aligned along the $z$ axis.
In the PN approximation, the leading-order gravitational waveform is
determined by the $l=2$, $m=2$ mode $U_{2,2,}$~\cite{Blanchet:2006zz}.
In terms of the PN parameter $x=(M\omega)^{2/3}$, where $\omega$ is the 
orbital frequency and $M$ is the total mass of the binary, it is given by 
\begin{equation}
U_{2,2} = -8\sqrt{\frac{2\pi}{5}} \eta M x \exp[i x^{-5/2}/(16\eta)]
+ O(c^{-2}) \, ,
\label{eq:U22}
\end{equation}
We have defined the symmetric mass ratio, $\eta = m_1 m_2/M^2$, where $m_1$ 
and $m_2$ are the masses of the two components of the binary.
The errors of $O(c^{-2})$ are again relative errors, and they arise in both 
the amplitude and the phase.
The time derivative of the mode can be computed via the chain rule 
\begin{equation}
\dot U_{lm} = \frac{dU_{lm}}{dx} \dot x \, .
\end{equation}
The quantity $\dot x$ also has a PN expansion which is given by
\begin{equation}
\dot x = \frac{-\mathcal L}{dE/dx} = \frac{64 \eta}{5M} x^5 + O(c^{-2})
\label{eq:xdot}
\end{equation}
where $\mathcal L$ is the GW luminosity.
Combining these results, we find that
\begin{equation}
\dot U_{2,2} = 16i \sqrt{\frac{2\pi}{5}} \eta x^{5/2} 
\exp[i x^{-5/2}/(16\eta)] + O(c^{-2}) \, .
\label{eq:U22dot}
\end{equation}
We will use both Eqs.~\eqref{eq:U22} and~\eqref{eq:U22dot} in our calculation
of the PN-expanded spin memory.

\subsection{\label{subsec:Bondi} Bondi framework}

The Bondi framework uses a set of coordinates $(u,r,\theta^A)$, with $A=1,2$.
The coordinates have the following interpretations: $u$ is a retarded time, 
$r$ is an affine parameter along outgoing null rays, and $\theta^A$ are 
arbitrary coordinates on the 2-sphere.
A set of boundary conditions are imposed upon the components
of the metric, so that it can be written in the form
\begin{align}
ds^2 = & -(1-2mr^{-1})du^2 +2du dr + D^B C_{AB} d\theta^A du \nonumber \\
& + r^2 (h_{AB} + r^{-1} C_{AB} )d\theta^A d\theta^B + \ldots \, ,
\end{align}
where the ellipsis denotes higher-order terms in the series in $1/r$.
The expression $h_{AB}$ is the metric on a round 2-sphere, $D_A$ is the 
covariant derivative compatible with $h_{AB}$, $C_{AB}(u,\theta^C)$ is the 
shear tensor, and $m(u,\theta^A)$ is the Bondi mass aspect.
The tensor $C_{AB}$ is symmetric and trace-free.
In Bondi gauge, 
\begin{equation}
N_{AB} = \partial_u C_{AB}
\label{eq:NewsEq}
\end{equation}
is the news tensor, which vanishes in stationary spacetimes.
The Einstein equations prescribe the evolution of $m$ and a higher-order
term in the metric $N_A(u,\theta^A)$.
Additionally, the evolution of $C_{AB}$ is set via Eq.~\eqref{eq:NewsEq}, but 
the news can be chosen freely.

Let us expand $C_{AB}$ in tensor harmonics
\begin{equation}
C_{AB} = \sum_{lm} (C_{(e),lm} T^{(e),lm}_{AB} + C_{(b),lm} T^{(b),lm}_{AB})
\, .
\label{eq:CABharmonics}
\end{equation}
The tensor $C_{AB}$ contains information about the gravitational-wave
strain; thus, one might expect there to be some relationship between it 
and the strain $h_{ij}^{\mathrm TT}$ in the PN approach, in a regime
in which both expansions are valid.
The easiest way to find the relationship is to compare the leading-order
part of the Riemann tensors in a series in $1/r$ in the linearized 
approximation.
In terms of the strain $h_{ij}^{\mathrm TT}$, the coordinate components of the
Riemann (or, equivalently at this order, Weyl) tensor are given by
\begin{equation}
R_{titj} = -\frac 12 \ddot h_{ij}^{\mathrm TT} + O(1/r^2) \, ,
\end{equation}
(see, e.g.,~\cite{Flanagan:2005yc}), whereas for the Bondi framework, the
leading-order components are given by
\begin{equation}
R_{uAuB} = -\frac 1{2r} \ddot C_{AB} + O(1/r^2)
\end{equation}
(see, e.g.,~\cite{Flanagan:2015pxa}).
Consider now the projection of the Riemann tensor into the orthogonal basis
vectors that are orthogonal to the propagation direction of the gravitational
waves, $\vec e_{\hat r} = \vec \partial_r$, 
\begin{equation}
\vec e_{\hat 0} = \vec \partial_u \, , \qquad 
\vec e_{\hat A} = \frac 1r \partial_A \, .
\end{equation}
After transforming between the Bondi and inertial Minkowski coordinates, 
we are then able to relate the coefficients $C_{(e),lm}$ and $C_{(b),lm}$ 
to the radiative multipole moments $U_{lm}$ and $V_{lm}$.
They satisfy the relationship
\begin{equation}
\ddot C_{(e),lm} = \ddot U_{lm} \, , \qquad 
\ddot C_{(b),lm} = \ddot V_{lm} \, .
\end{equation}
By fixing the constants of integration, we can construct a Bondi frame in
which 
\begin{equation}
C_{(e),lm} = U_{lm} \, , \qquad C_{(b),lm} = V_{lm} \, .
\label{eq:CABmoments}
\end{equation}
In such a frame, we can then use the Bondi approach in the regime of validity 
of the PN approximation, which will allow us to compute the displacement and
spin memories in this context.

\section{\label{sec:disp_memory} Multipolar expansion of the displacement 
memory within the Bondi framework}

We first derive a general expression for the displacement memory in terms
of a set of radiative multipole moments; we then specialize the expression
for nonspinning compact binaries in quasicircular orbits in the PN 
approximation.

\subsection{Displacement memory in terms of radiative multipole moments}

In the Bondi framework, the total memory can be computed from the 
expression
\begin{equation}
\mathcal D \Delta\Phi = 8 \Delta m + \int du (32\pi r^2 T_{uu} + N_{AB}N^{AB})
\, ,
\label{eq:memory}
\end{equation}
where the limits of integration are formally $u\rightarrow\pm\infty$ 
(see, e.g.,~\cite{Flanagan:2015pxa}).
In Eq.~\eqref{eq:memory}, we have defined several quantities:
the operator $\mathcal D = D^2(D^2+2)$, the field $\Delta \Phi$ which is
related to the change in the electric-parity part of the shear tensor in
the Bondi framework
\begin{equation}
\Delta C_{AB} = \frac 12 (2D_A D_B - h_{AB} D^2) \Delta\Phi \, , 
\end{equation}
and the energy per unit solid angle in matter fields, $T_{uu}$, which scales 
as $r^{-2}$.
Note that the quantity $\Delta C_{AB}$ is strictly electric parity for most
astrophysical sources of gravitational waves 
(see, e.g.,~\cite{Madler:2016ggp}), while the tensor $C_{AB}$ generically has 
electric- and magnetic-parity parts [see Eq.~\eqref{eq:CABphiPsi}].

To solve for $\Delta\Phi$, we must invert the operator $\mathcal D$.
This operator has a nontrivial kernel (the $l=0$ and $l=1$ spherical 
harmonics), and its inverse is not defined on these elements of the
kernel.
However, if we define a projector $\mathcal P$ (the exact expression for which 
is not particularly important) that removes these first two harmonics and 
apply it to the right-hand side of Eq.~\eqref{eq:memory}, then we can solve
for $\Delta\Phi$.
The result is
\begin{equation}
\Delta\Phi = \mathcal D^{-1} \mathcal P\left[8\Delta m + 
\int du (32\pi r^2 T_{uu} + N_{AB}N^{AB}) \right] \, .
\end{equation}
In the PN approximation, only the nonlinear memory is typically computed.
Thus, we will treat the contribution of just the last term,
\begin{equation}
\Delta\Phi = \mathcal D^{-1} \mathcal P\int_{-\infty}^{u_f} du N_{AB}N^{AB} 
\, .
\label{eq:NullMemory}
\end{equation}
As is also common in the PN approximation, we truncate the integral
at a finite retarded time $u_f$, rather than taking the formal limit 
as $u\rightarrow\infty$.
We will now compute the multipole moments of Eq.~\eqref{eq:NullMemory},
which are given by
\begin{equation}
\Delta\Phi_{lm} = \frac{(l-2)!}{(l+2)!}\int d^2\Omega \int_{-\infty}^{u_f} 
du N_{AB} N^{AB} \bar Y_{lm} \, ,
\label{eq:NullMemoryMoments}
\end{equation}
where $l\geq 2$.

Before we expand the news tensor in radiative multipoles, it will be useful
to introduce some notation for the integral of three spin-weighted spherical
harmonics over $S^2$,
\begin{equation}
\int d^2\Omega ({}_{s'}Y_{l'm'}) ({}_{s''}Y_{l''m''}) ({}_{s}\bar Y_{lm}) \, .
\end{equation}
These integrals have been performed in several different sources, with nearly
as many different conventions for the definitions and normalizations of the
spin-weighted spherical harmonics.
We use a notation most similar to that in~\cite{Beyer:2013loa} for these
integrals, but we use the conventions for the spin-weighted spherical harmonics
in~\cite{Ruiz:2007yx} (which we give in Appendix~\ref{sec:harmonic_relations} 
for completeness).
These integrals are only nonzero when $s=s'+s''$, $m=m'+m''$, and $l$ 
is in the range 
$\Lambda = \{\max(|l'-l''|,|m'+m''|,|s'+s''|),\ldots,l'+l''-1,l'+l''\}$.
The integral can be expressed in terms of Clebsch-Gordon coefficients, and we
will denote it as
\begin{align}
& \int d^2\Omega ({}_{s'}Y_{l'm'}) ({}_{s''}Y_{l''m''}) 
({}_{s'+s''}\bar Y_{lm'+m''}) \equiv \nonumber \\
& \mathcal C_l(s',l',m';s'',l'',m'') \, ,
\label{eq:3sYlm}
\end{align}
where we have defined 
\begin{align}
& \mathcal C_l(s',l',m';s'',l'',m'') = (-1)^{l+l'+l''}
\sqrt{\frac{(2l'+1)(2l''+1)}{4\pi(2l+1)}}
\nonumber \\
& \times \langle l',s';l'',s''| l,s'+s''\rangle 
\langle l',m';l'',m''| l,m'+m''\rangle \, .
\end{align}
We have used a bra-ket notation for the Clebsch-Gordon coefficients.
Two useful properties of the coefficients 
$\mathcal C_l(s',l',m';s'',l'',m'')$ that we will frequently use are
\begin{subequations}
\begin{align}
\mathcal C_l(s',l',m';s'',l'',m'') = & (-1)^{l+l'+l''} \nonumber \\
& \times \mathcal C_l(-s',l',m';-s'',l'',m'') \, , \\
\mathcal C_l(s',l',m';s'',l'',m'') = & (-1)^{l+l'+l''} \nonumber \\
& \times C_l(s',l',-m';s'',l'',-m'') \, ,
\end{align}
\label{eq:Alprops}%
\end{subequations}
where we have restricted to integer spin weights $s'$ and $s''$.

We can then express Eq.~\eqref{eq:NullMemoryMoments} in terms of the radiative
moments $U_{lm}$ and $V_{lm}$ by substituting Eqs.~\eqref{eq:CABharmonics} and 
\eqref{eq:CABmoments} into Eq.~\eqref{eq:NullMemory} and by using 
Eqs.~\eqref{eq:TABharmonics},~\eqref{eq:Rank2Spin2Ylm}, and~\eqref{eq:Alprops}.
We find that the moments $\Delta\Phi_{lm}$ can be written as
\begin{align}
\Delta\Phi_{lm} = & \frac 12 \frac{(l-2)!}{(l+2)!}
\sum_{l',l'',m',m''} \mathcal C_l(-2,l',m';2,l'',m'') \nonumber \\
& \times \int_{-\infty}^{u_f} du \{ 2i[1-(-1)^{l+l'+l''}]
\dot U_{l'm'} \dot V_{l''m''} \nonumber \\
& + [1+(-1)^{l+l'+l''}] (\dot U_{l'm'} \dot U_{l''m''} 
+ \dot V_{l'm'} \dot V_{l''m''}) \} \, .
\label{eq:NullMemGenInt}
\end{align}
The expression above requires that $l,l',l''\geq 2$.
Only terms with $l\in\Lambda$ are nonzero.
In addition, just the terms in the sum with $m'+m''=m$ contribute.
The expression in Eq.~\eqref{eq:NullMemGenInt} involves an infinite double 
sum over the moments, but in the PN expansion, only a very small number 
contribute at the leading order.

\subsection{\label{subsec:MemPN} Displacement memory for compact binaries in 
the PN approximation}

When computing the displacement memory for nonspinning compact binaries, it is 
convenient to choose a coordinate system in which the orbital angular momentum 
of the binary lies along the $z$ axis.
Favata \cite{Favata:2008yd}, for example, has shown that the $m=0$ modes of 
the memory observable $\Delta\Phi_{l0}$ are the only modes that enter at the
leading PN order and that are nonoscillatory (the modes of $\Delta\Phi_{lm}$ 
with $m\neq 0$ introduce oscillatory corrections to the waveform that begin 
at a relative 2.5PN order).
Thus, we will specialize to the $m=0$ modes of $\Delta\Phi_{lm}$.

For simplicity, we just compute the leading-order memory effect,
which is encoded in the $\Delta\Phi_{2,0}$ and $\Delta\Phi_{4,0}$ modes
(not surprisingly, it is only $U_{2,2}$ that is needed) so that $\Delta\Phi$
is given by 
\begin{align}
\Delta \Phi = & \frac 1{15120\sqrt{\pi}} (180\sqrt{5} Y_{2,0} + Y_{4,0})
\int^{u_f}_{-\infty} du |\dot U_{2,2}|^2 \nonumber \\
& + O(c^{-2}) \, ,
\label{eq:MemModes2040}%
\end{align}
where $u_f$ is the retarded time at which the integral is truncated.
Computing the memory reduces to a question of how to evaluate the integral
in Eq.~\eqref{eq:MemModes2040}.
Favata~\cite{Favata:2008yd} showed that one way to do this is to transform
the integral with respect to $u$ to one with respect to $x$ by dividing by 
$\dot x$.
Substituting Eq.~\eqref{eq:U22dot} into~\eqref{eq:MemModes2040}
and using Eq.~\eqref{eq:xdot}, we find that
\begin{equation}
\Delta \Phi =  \frac{\sqrt{\pi}}{1890} M\eta x_f 
(180\sqrt{5} Y_{2,0} + Y_{4,0}) + O(c^{-2}) \, .
\end{equation}
The quantity $x_f$ is given by $x$ evaluated at $u_f$.
Next, we reconstruct $\Delta C_{AB}$ from $\Delta\Phi$ to find
\begin{equation}
\Delta C_{AB} = \frac{\sqrt{5\pi}}{315} M \eta x_f (60\sqrt{3} T^{(e),2,0}_{AB} 
+  T^{(e),4,0}_{AB}) + O(c^{-2}) \, .
\end{equation}
Thus, the strain $h=h_+ - i h_\times$ is given just by $h_+$ to this order
and has the form
\begin{equation}
h_+^{\mathrm{mem}} =  \frac{\sqrt{5\pi} }{315\sqrt 2 r} M \eta x_f 
(60\sqrt{3} {}_{-2}Y_{2,0} + {}_{-2}Y_{4,0}) + O(c^{-2}) \, .
\end{equation}
From this, we recover the expression for the memory
\begin{equation}
h_+^{\mathrm{mem}} = \frac 1{48r} M \eta x_f \sin^2\theta(17+\cos^2\theta) 
+ O(c^{-2})\, ,
\end{equation}
which was first computed in~\cite{Wiseman1991} (though which differs
by a known factor of two~\cite{Arun:2004ff}).
We take this result as an indication that we can correctly incorporate
leading-order results in the PN approximation into a specific Bondi frame.
This ability will be necessary to compute the leading-order part of the spin
memory in the next section.

\section{\label{sec:spin_memory} Multipolar expansion of the spin memory within
the Bondi framework}

We begin this section by deriving an expression for the nonlinear null
spin memory in terms of a set of radiative multipole moments.
We next compute the spin memory in the PN approximation for nonspinning, 
quasicircular compact binaries, and we conclude this section by articulating
the relationship between the rate of accumulation of the spin memory and 
2.5PN nonhereditary, nonoscillatory, and nonlinear term in the gravitational
waveform.
The steps involved in the calculation of the spin memory are quite similar to
those in the computation of the memory in the previous section.

\subsection{Spin memory in terms of radiative moments}

The shear tensor $C_{AB}$ can be written as a sum of two terms
\begin{equation}
C_{AB} = \frac 12 (2 D_A D_B - h_{AB} D^2) \Phi 
+ \epsilon_{C(A} D_{B)} D^C \Psi \, .
\label{eq:CABphiPsi}
\end{equation}
The functions $\Phi$ and $\Psi$ are smooth functions of the coordinates
$(u,\theta^A)$.
The spin memory observable was defined to be 
\begin{equation}
\Delta\Sigma \equiv \int du \Psi \, ,
\end{equation}
where the limits of the integrand are $u\rightarrow\pm\infty$
(see, e.g.,~\cite{Flanagan:2015pxa}).

The quantity $\Delta\Sigma$ is determined by changes in the flux of
angular momentum per unit solid angle in gravitational waves and matter
and changes in the curl of a quantity $\Delta\hat N_A$ that is a 
generalization of the spin of the system (the superspin charges).
The relationship is
\begin{align}
D^2 \mathcal D \Delta\Sigma = & 8 \epsilon^{AB} D_B \Delta \hat N_A + 
64\pi r^2 \int du \epsilon^{AB} D_B T_{uA} + \nonumber \\
& \int du \epsilon^{AD} D_D (3N_{AB} D_C C^{BC} - 
3C_{AB} D_C N^{BC} \nonumber \\
& - N^{BC} D_B C_{AC} + C^{BC} D_B N_{AC} ) \, ,
\label{eq:SpinMemory}
\end{align}
where $T_{uA}$ is the angular momentum per unit solid angle in matter fields,
which scales with $r$ as $r^{-2}$.
We will focus on just the vacuum, null part of the spin memory, which is just
the last two lines of Eq.~\eqref{eq:SpinMemory}, because of the close analogy 
to computing just the null, vacuum part of the displacement memory in 
Sec.~\ref{sec:disp_memory}.
In doing so, we will again make use of the operator $\mathcal P$ that 
projects out the $l=0$ and $l=1$ spherical harmonics.
The nonlinear, null, spin memory can be calculated from
\begin{align}
D^2 \mathcal D \Delta\Sigma = & \mathcal P \int du \epsilon^{AD} D_D 
( C^{BC} D_B N_{AC} - N^{BC} D_B C_{AC} \nonumber \\
& + 3N_{AB} D_C C^{BC} - 3C_{AB} D_C N^{BC})  \, .
\label{eq:NullSpinMemory}
\end{align}

For computing this spin memory, it will be useful to expand $\Delta\Sigma$
in scalar spherical harmonics so as to obtain the moments $\Delta\Sigma_{lm}$.
Integrating by parts, one can then express the right-hand side as the moments
of a vector field with respect to the magnetic-parity vector 
harmonics.\footnote{Strictly speaking, the extended BMS algebra of Barnich and
Troessaert is spanned by a Virasoro basis rather than one of smooth vector 
fields. 
However, more recent work of Campiglia and 
Laddha~\cite{Campiglia:2014yka, Campiglia:2015yka} suggested
that the generalized BMS group that contains all smooth diffeomorphisms of the 
2-sphere rather than just the Lorentz symmetries could be an equally valid 
extension of the BMS group.
It is following Campiglia and Laddha that we compute the multipole moments
of $\Delta\Sigma$ for the spin memory with respect to the smooth 
magnetic-parity vector harmonics.}
Because $C_{AB}$ and $N_{AB}$ are symmetric trace-free tensors, then 
the terms $C^{BC} D_B N_{AC}$ and $N^{BC} D_B C_{AC}$ only involve the
symmetric trace-free part of $D_B N_{AC}$ and $D_B C_{AC}$.
By writing $C_{AB}$ in terms of the multipole moments $U_{lm}$ and $V_{lm}$,
expanding its derivatives, and using some of the relations in 
Appendix~\ref{sec:harmonic_relations}, we can reduce the angular integral
to the product of three spin-weighted spherical harmonics, as in 
Eq.~\eqref{eq:3sYlm}.
Taking these facts into account, we find
\begin{widetext}
\begin{align}
\Delta\Sigma_{lm} = & \frac{1}{4\sqrt{l(l+1)}} \frac{(l-2)!}{(l+2)!}
\sum_{l',l'',m',m''} 
[3\sqrt{(l'-1)(l'+2)} \mathcal C_l(-1,l',m';2,l'',m'')
+\sqrt{(l''-2)(l''+3)} 
\nonumber \\
& \times \mathcal C_l(-2,l',m';3,l'',m'')]
\int_{-\infty}^{u_f} du \{ 
i[1-(-1)^{l+l'+l''}] (U_{l'm'} \dot U_{l''m''} 
 - \dot U_{l'm'} U_{l''m''} + V_{l'm'} \dot V_{l''m''} 
- \dot V_{l'm'} V_{l''m''})
\nonumber \\
& -[1+(-1)^{l+l'+l''}]
(U_{l'm'} \dot V_{l''m''} + \dot V_{l'm'}U_{l''m''} 
 - \dot U_{l'm'} V_{l''m''} - V_{l'm'} \dot U_{l''m''} ) \} \, .
\label{eq:NullSpinMemMom}
\end{align}
\end{widetext}
The expression above requires $l,l',l''\geq 2$.
As with the displacement memory, only terms with $l\in\Lambda$ are nonzero
and only those with $m'+m''=m$ contribute to the sum.

The $l=1$ moments of the right-hand side of Eq.~\eqref{eq:NullSpinMemory}
[without the projector, or similarly for Eq.~\eqref{eq:NullSpinMemMom}]
contain information about the angular momentum radiated, which was derived 
in~\cite{Thorne1980} through angular-momentum balance in the center-of-mass
frame of a PN source.
As a check of our expression for the spin memory, we show in 
Appendix~\ref{sec:AngMomFlux} that we can derive the expression for the flux 
of angular momentum in terms of the radiative multipole moments of the
gravitational wave strain given in~\cite{Thorne1980}.
Next, we turn to the spin memory of compact binaries.

\subsection{Spin memory for compact binaries in the PN approximation}

The expression~\eqref{eq:NullSpinMemMom} greatly simplifies in the PN
approximation for nonspinning, quasicircular binaries.
We will consider the same set up as in Sec.~\ref{subsec:MemPN}:
a binary with orbital angular momentum in the $z$ direction.
At leading order in the PN approximation, the relevant multipoles will be
just $U_{2,2}$ and $\dot U_{2,-2}$ (plus their complex conjugates).
Arguments similar to those in Sec.~\ref{subsec:MemPN} suggest that 
the $m=0$ modes of $\Sigma_{lm}$ will be the leading-order effect,
whereas all the $m\neq 0$ modes will be relative 2.5PN corrections.
The relevant moment of the spin memory will be just the $l=3$,
$m=0$ mode, whereas all other modes will be higher-PN quantities.
Thus, $\Delta \Sigma$ is given by 
\begin{equation}
\Delta \Sigma =  \frac 1{80\sqrt{7\pi}} Y_{3,0} \int^{u_f}_{-\infty} du 
\Im ( \bar U_{2,2} \dot U_{2,2}) + O(c^{-2}) \, ,
\label{eq:SpinMemModes30}%
\end{equation}
where $\Im()$ denotes the imaginary part.
We will evaluate this integral by transforming to the coordinate $x$.
However, we will now assume that as $u\rightarrow -\infty$, $x$ 
approaches a small, but nonzero value $x_{-\infty}$ (namely, the binary
formed at a finite separation in the past).
We then find that $\Delta\Sigma_{3,0}$ takes the form
\begin{equation}
\Delta \Sigma = \frac 1{10} \sqrt{\frac{\pi}{7}} \eta M^2 
(x_f^{-1/2} - x_{-\infty}^{-1/2}) Y_{3,0} + O(c^{-2}) \, .
\end{equation}
The $u$ integral of the magnetic-parity part of $C_{AB}$ associated
with the spin memory takes the form
\begin{align}
\epsilon_{C(A} D_{B)} D^C \Delta\Sigma = & 
 \sqrt{\frac{3\pi}{35}} \eta M^2 (x_f^{-1/2} - x_{-\infty}^{-1/2}) 
T^{(b),3,0}_{AB} \nonumber \\
& + O(c^{-2}) \, .
\end{align}
Thus, we can infer that the contribution of the spin memory to the $u$ 
integral of the strain enters into the cross component
\begin{align}
\int_{-\infty}^{u_f} du h_\times^{\mathrm{smm}} 
= & \frac 1{r} \sqrt{\frac{3\pi}{70}} 
\eta M^2 (x_f^{-1/2} - x_{-\infty}^{-1/2}) {}_{-2}Y_{3,0} \nonumber \\
& + O(c^{-2}) \, .
\end{align}
We have labeled this contribution by ``smm'' for reasons that we will 
describe in Sec.~\ref{subsec:SpinMemMode}.
Inserting the explicit expression for the spin-weighted spherical harmonic
we find that
\begin{align}
\int_{-\infty}^{U_R} du h_\times^{\mathrm{smm}}
= & \frac{3}{8 r} M^2 \eta 
(x_{f}^{-1/2} - x_{-\infty}^{-1/2}) \sin^2\theta \cos\theta \nonumber \\
& + O(c^{-2}) \, .
\label{eq:SpinMemLeadingPN}
\end{align}
Note that if $x_{-\infty}$ were allowed to go to zero (having an infinitely 
separated binary), the spin memory observable would diverge in the infinite 
past; however, we argue in the next part that such a limit is not a physically
reasonble one in this context.

\subsection{\label{subsec:SpinMemFreq} Discussion of the frequency dependence
of the spin memory}

The frequency dependence of the spin memory effect might initially appear
somewhat problematic, because it scales with a negative power of frequency 
and would diverge if the frequency goes to zero (i.e., if the components of 
the binary are separated by an infinite distance).
Simply having a quantity scale with a negative power of frequency is not 
necessarily an issue; it might indicate that there is a well-defined
physical reason for the quantity to be infinite in the limit of infinite
separation.
For example, the leading-order gravitational-wave phase for compact binaries 
scales as $x^{-5/2}$ [see Eq.~\eqref{eq:U22}], but this simply reflects the 
fact that a binary must undergo an infinite number of cycles (over an 
infinitely long time) to inspiral from an infinitely distant separation.
Given the finite age of the universe, the binaries of observational relevance
are those that are formed at a finite separation, undergo a finite number of
cycles, and merge within a Hubble time.
The conclusion to draw from this example is not to abandon the 
gravitational-wave phase as a physical observable, but rather to restrict
the use of this observable to systems for which it is a finite quantity
(namely, binaries formed at a finite separation).

The angular momentum of a binary is another example of such a quantity.
In the Newtonian limit, it is given by $\eta M^2 \sqrt{r/M} \sim x^{-1/2}$;
therefore, it formally diverges for a binary with its components at an 
infinite separation.
Because it has been proven in the PN approximation that the angular momentum 
carried by the gravitational waves is equivalent to the change in the angular 
momentum of the binary (i.e., that the balance equations hold; see, 
e.g.,~\cite{Blanchet:1996vx}), then it follows that the radiated angular 
momentum will be equal at leading order to the change in the Newtonian angular 
momentum as the binary's separation decreases.
Thus, the leading-order change in angular momentum from an inspiraling binary 
system is another example of a quantity that is well defined physically, but 
which should be restricted to the class of systems for which it is finite.

The spin memory is roughly the ``higher multipole moments of the change in 
the radiated angular momentum'' for compact binaries.
In fact, the same combination of the multipole moments that enter into 
the leading-order spin memory effect also appear in the leading change in 
the radiated angular momentum.
This, in turn, gives the spin memory the same time and frequency dependencies
as the change in angular momentum have.
Therefore, it seems natural to treat the spin memory with the same set of 
physical constraints that one would for the change angular momentum: namely, 
that one should not consider the spin memory in the limit a binary with an 
infinite separation of its components.
In addition, because any actual compact binary in our Universe has a finite 
initial separation and a limited change in frequencies, this is not a 
significant physical restriction on the class of solutions that we consider.

\subsection{\label{subsec:SpinMemMode} Relation to nonhereditary, 
nonoscillatory contributions to compact-binary waveforms}

In~\cite{Arun:2004ff}, the polarizations of the gravitational waveform were
computed for compact binaries in quasicircular orbits at 2.5PN order.
At this order, a new contribution to the cross polarization of the waveform
was discovered that was nonlinear, nonoscillatory, but also not 
hereditary (in the sense that it does not involve an integral over
the multipole moments over all retarded times up to the given time; rather, 
it depends on the value of the moments at that given retarded time).
It appeared in the radiative current octopole moment as a result of a 
nonlinear interaction between the mass quadrupole moments of the source
and also the mass quadupole and spin dipole.
The interpretation of this term was not completely understood.
For example, whether this term would give rise to a constant offset between
early and late times had not been known (see the discussion 
in~\cite{Favata:2008yd} for more details).\footnote{Recent work by 
M\"adler and Winicour~\cite{Madler:2016ggp} suggests that a displacement 
memory effect with a magnetic-parity sky pattern only arises as a result of 
incoming gravitational waves from past null infinity; it would, therefore, be 
primordial in origin.
However, PN theory specifically assumes stationarity in the past so as to
eliminate incoming radiation from past null infinity; therefore, one would
generally expect that this contribution to the cross polarization---which
has a magnetic-parity sky pattern---would be identically zero at early 
and late times.
Thus, while this argues against the nonhereditary, nonoscillatory term 
as resulting in a constant offset between early and late times, it does not 
explain what its origin and interpretation might be.}

By differentiating Eq.~\eqref{eq:SpinMemLeadingPN} with respect to $u$,
we find that
\begin{equation}
h_\times^{\mathrm{smm}} = -\frac{12 M\eta^2}{5r} x^{7/2} 
\sin^2\theta \cos\theta + O(c^{-2}) \, ,
\end{equation}
which is precisely the expression given in~\cite{Arun:2004ff}, after accounting
for the relative minus sign in our conventions for the gravitational-wave
polarizations.
It will also be useful to express this quantity in terms of the radiative
mass quadrupole moments $U_{2,2}$.
The result of this is
\begin{equation}
h_\times^{\mathrm{smm}} = \frac{3}{64\pi r} \Im(\bar U_{2,2} \dot U_{2,2}) 
\sin^2\theta \cos\theta \, .
\label{eq:SpinMemMode}
\end{equation}
Because this part of $h_\times$ is intimately connected to the spin memory,
we will call $h_\times^{\mathrm{smm}}$ the ``spin memory mode'' (and which
is why we have labeled it with the abbreviation ``smm'').

In~\cite{Favata:2008yd}, there is also a general expression for the 
term $h_\times^{\mathrm{smm}}$ in terms of the source multipole moments.
By integrating the expression~\eqref{eq:SpinMemModes30} by parts, 
differentiating the resulting expression, and using
the relationships between the radiative and source multipole moments in
Eq.~\eqref{eq:RadiativeToSource} we can recover the expression quadratic in
the source mass quadrupole moments in Eq.~(5.6) of~\cite{Favata:2008yd}.
However, there is an additional contribution to that expression involving
the interaction of the spin dipole and mass quadrupole, which is not included
in the $u$ derivative of the spin memory.
It is interesting to note that this additional term does not contribute for
the compact binaries that are the focus of this paper.

\section{\label{sec:detection} Prospects for detection of the spin memory
mode}

We primarily focus on the detectability of the spin memory mode by
ground-based gravitational-wave detectors; however, we will also comment 
on detecting the spin memory from other sources and by different types
of detectors.

\subsection{Compact binaries with ground-based detectors}

We first describe how we compute the approximate spin memory mode using
the SXS:BBH:0305 waveform from the SXS waveform catalog~\cite{SXS:catalog}.
We next describe  how we compute the SNR for such waveforms.
Finally, we sketch how it could be possible to coherently combine multiple
subthreshold observations of the spin memory mode to make a detection
of the effect from multiple compact-binary mergers.

\subsubsection{Construction of an approximate spin memory mode}

Although the spin memory accumulates over a long timescale during the inspiral
of a binary, it does not seem possible to measure this aspect of it with 
ground-based interferometers; rather, it will be the rate of change of the
spin memory near the binary's merger that will generate a signal that can be
measured by such detectors.
This part of the binary's evolution, however, is not treated accurately by the 
PN approximation.
Numerical relativity is capable of simulating this portion of the waveform,
in principle, using Cauchy-characteristic extraction; however, such waveforms
are not publicly available and those without it are known to not represent
the relevant modes accurately (see, e.g., the technical caveats 
in~\cite{SXS:catalog}).
Thus, a method to estimate the spin memory mode will be necessary to 
determine whether it is observable or not.

To do this, we will use the expression for the leading-order spin memory
mode in Eq.~\eqref{eq:SpinMemMode}, but we will substitute in the 
$U_{2,2}$ and $U_{2,-2}$ modes from the simulation SXS:BBH:0305.
The resulting spin memory mode is shown in Appendix~\ref{sec:SpinMemNR},
where it is directly compared with the corresponding mode from 
numerical relativity simulations.
The spirit of this approach is similar to that of~\cite{Favata:2009ii},
in which the displacement memory is computed from an expression of the form
\begin{equation}
h_+^{\mathrm{mem}} = \frac{1}{384\pi r} \sin^2\theta (17+\cos^2\theta)
\int du |\dot U_{2,2}|^2 \, ,
\label{eq:DispMemMode}
\end{equation}
for radiative multipoles $U_{2,2}$ that were chosen to fit the 
numerical-relativity waveform modes.
When we compute the SNR for the displacement memory modes in the next 
part of this section, we will use the expression~\eqref{eq:DispMemMode},
with the $\dot U_{2,2}$ mode from the numerical simulation.

\subsubsection{SNR calculations from individual compact-binary coalescences}

\begin{table}
\caption{SNR of dominant ($l=2$, $m=2$) mode, displacement memory mode 
($l=2,4$, $m=0$), and spin memory mode ($l=3$, $m=0$) in advanced LIGO in
the zero-detuning, high-power configuration and in the Einstein Telescope in 
the ET-B design.
The source is a compact binary with the masses, spins, luminosity distance,
and inclination of the GW150914 event; however, the sky location, reference
phase, and polarization are chosen to be maximal for the respective 
modes (see the text for more detail).
Just one digit of accuracy is given.}
\begin{tabular}{ccc}
\hline
\hline 
Mode & SNR in LIGO & SNR in ET \\
\hline
$l=2$, $m=2$ & $\sim 1\times 10^2$ & $\sim 5\times 10^3$ \\
$l=2,4$, $m=0$ & $\sim 1$ & $\sim 30$ \\
$l=3$, $m=0$ & $\sim 0.03$ & $\sim 1$ \\
\hline
\hline
\end{tabular}
\label{tab:SNRtable}
\end{table}

To estimate the circumstances in which the spin memory mode will be detectable,
we perform calculations of the SNR.
The square of the SNR is defined by
\begin{equation}
\rho^2 = 4 \int_0^\infty df \frac{|\tilde h(f)|^2}{S_n(f)} \, ,
\end{equation}
where $S_n(f)$ is the noise power-spectral density of the detector and 
$\tilde h(f)$ is the Fourier transform of the detector response
\begin{equation}
h(t) = F_+ h_+ + F_\times h_\times \, ,
\end{equation}
where $F_+$ and $F_\times$ are the antenna response patterns of the detectors
to the plus and cross polarizations of the waveform.
We will compute the SNR in the zero-detuning, high-power configuration of
advanced LIGO and the ET-B design of the Einstein Telescope.
Fits of the power-spectral densities of these two detectors are given 
in~\cite{Ajith:2011ec} and~\cite{Regimbau:2012ir}, respectively.

For the waveforms, we consider signals comprised of just the dominant,
quadrupole mode ($l=2$, $m=2$), the displacement memory mode ($l=2,4$, $m=0$),
and the spin memory mode ($l=3$, $m=0$).
In the SNR calculation for advanced LIGO, the $l=2$, $m=2$ mode comes from 
the simulation SXS:BBH:0305, which was the simulation used in the 
paper~\cite{Abbott:2016blz} with the most likely parameters of the GW150914 
event.
The waveform in the SXS catalog has the mass and distance normalized
out of the waveform; therefore, we scale it by the most likely luminosity 
distance ($D_L=410$Mpc) and the total source-frame mass of the binary 
($65M_\odot$) at a redshift of $z=0.09$.
Similarly to~\cite{Lasky:2016knh}, we also assume that the inclination of the
binary is given by $\theta \sim 7\pi/9$, which is consistent with the 
posteriors for the GW150914 event.
This inclination angle gives rise to a nearly maximal spin memory signal for 
this event and a strong (though not maximal) displacement memory signal and 
dominant, quarupole signal.
For the sky position, polarization, and phase at coalescence, we choose these
values to maximize the SNR for the respective types of signals.
In the SNR calculation for the Einstein Telescope, we use the $l=2$, $m=2$
mode from the PhenomC waveform model~\cite{Santamaria:2010yb}, because this
model covers lower gravitational-wave frequencies that are not captured in the 
numerical relativity simulation.\footnote{Although it is necessary to use a
waveform model that includes frequencies lower than those contained in the 
numerical relativity waveforms to accurately compute the SNR from the 
$l=2$, $m=2$ waveform in ET-B, we have checked that the SNR of the 
displacement memory and spin memory modes are nearly identical when computed 
with the numerical relativity waveform and the PhenomC waveform. 
For these memory modes, the contribution to the SNR from times earlier than 
the starting time of the numerical relativity waveform is negligible.}
The parameters of the PhenomC waveform are chosen to be the same as those for 
the numerical waveform.

Our results are summarized in Table~\ref{tab:SNRtable}.
For individual binary mergers, the spin memory mode will be too small to 
detect even for the Einstein Telescope.
However, it is of interest to note that the displacement memory mode can 
have a significant SNR for GW150914-like events in ET-B.

\subsubsection{Stacking multiple subthreshold spin-memory-mode observations}

Because detecting the spin memory from individual binary mergers is unlikely,
trying to coherently combine multiple subthreshold spin-memory-mode 
observations is the more promising route towards detection.
Coherent addition of the signals allows for the evidence to grow with the 
number of detections $N$ as $\sqrt{N}$~\cite{Lasky:2016knh}.
The strategy for detecting the spin memory would run similarly to that 
elaborated in~\cite{Lasky:2016knh} for the displacement memory modes, 
but there is one additional subtlety for which we must account.

To coherently add the displacement memory modes, one must determine the sign 
of the strain signal in the detector that is associated with the displacement
memory.
The sky pattern of the memory produces a strictly nonnegative strain, 
$h_+^{\mathrm{mem}}$; therefore, the sign in the detector's response arises
solely from the polarization angle $\psi$ in the detector's antenna
response function (i.e., the sign of the memory is determined just by the
sign of the antenna response function).
However, it is not possible to determine the polarization angle and the sign
of the antenna response function from a measurement of the leading 
$l=2$, $m=2$ modes of the waveform, because there is a degeneracy between the
reference phase (phase at coalescence) and the polarization under the shift
\begin{equation}
(\phi_c',\psi') = (\phi_c+\pi/2, \psi+\pi/2) \, .
\end{equation}
A shift of $\psi$ by $\pi/2$ changes the sign of the antenna response function
and thence the displacement memory signal in the detector.
Fortunately, this degeneracy can be broken using higher-order, spin-weighted
spherical-harmonic modes of the gravitational waveform.
Specifically, when the SNR in the combination of higher-order modes
\begin{equation}
\Delta h = \sum_{l,m} [h_{lm}(\phi_c,\psi;\vec\theta) - 
h_{lm}(\phi_c',\psi';\vec\theta)] ({}_{-2}Y_{lm})
\label{eq:DeltahDegen}
\end{equation}
is sufficiently high, it is possible to determine the sign of the 
detector's antenna response function [note that the term
$\vec\theta$ in Eq.~\eqref{eq:DeltahDegen} is used to indicate that the 
spherical-harmonic modes depend on other parameters that are not changed, and
also note that we use the convention of~\cite{Lasky:2016knh} in which 
quantities like the phase at coalescence and the antenna response pattern are
also included in the quantity $h_{lm}$ in Eq.~\eqref{eq:DeltahDegen}].

All these subtleties for detecting the displacement memory mode would still
apply to the spin memory mode; however, there is now one additional issue:
the spin memory mode changes sign under a change in the inclination angle
$\iota\rightarrow \pi-\iota$.
Therefore, to coherently add the spin memory signals, we would also need
to measure the inclination with sufficient accuracy to determine
the sign of the sky pattern of the spin memory mode.
Given the very high SNR of GW150914-like events in planned third-generation 
detectors like the Einstein Telescope, it is likely that this will be possible.
In addition, given the comparable SNRs of the spin memory mode in the Einstein 
Telescope and the displacement memory mode in advanced LIGO, it is plausible 
that on the order of one-hundred binary-black-hole mergers will need to be 
added coherently to detect the spin memory.
However, a more detailed investigation into the number of sources and the best 
measurement procedures needed to detect the spin memory mode is beyond the 
scope of this paper.
It will be left for future work.

\subsection{Discussion of other types of sources and detectors}

Because the spin memory is defined to be the change in the magnetic-parity 
part of the time-integrated gravitational-wave strain, it would make for a
more compelling detection if there were a detector (and source) that could
directly access this observable.
There is, in principle, an experiment that could measure this quantity 
directly: pulsar timing arrays.
The residuals depend upon the difference of the time integral of the 
gravitational-wave strain at Earth and at the pulsars in the timing array
(see, e.g.,~\cite{Lommen:2015gbz}).
For a gravitational-wave event within the galaxy, such as a galactic supernova,
the pulsars could surround the source of waves.
If so, they can directly extract the magnetic-parity part of the sky pattern 
of the gravitational-wave strain.
Given the infrequency of such supernovae and the uncertainties about the 
amount of angular momentum radiated in gravitational waves or neutrinos from 
such events, it is difficult to quantify the prospect of detecting such an 
effect.
But if there were a detection of displacement memory by pulsar timing arrays,
it would be of interest to search for any coincident spin memory effects 
as well.

The spin memory mode from individual, resolvable supermassive black-hole 
binaries is a potential source for space-based gravitational-wave 
interferometers, like the LISA mission~\cite{Audley:2017drz}.
Given the uncertainties on the rates of such mergers and the SNR of the
loudest events in the LISA band, it is more difficult to get a quantitative
estimate of the number of events needed to detect the spin memory mode with 
the LISA detector.
However, it might be of interest to explore this in more detail in future work.

\section{\label{sec:conclude} Conclusions}

In this paper, we computed the spin memory effect within the Bondi
framework in terms of a multipolar expansion of the gravitational-wave strain.
This allowed us to relate the radiative multipoles to source multipoles
via the PN approximation and compute the spin memory from specific sources
of gravitational waves.
We focused on nonspinning, quasicircular compact binaries, and we found that 
the spin memory has some interesting differences from the displacement memory.
Because the displacement memory grows like $x=(M\omega)^{2/3}$, it is largest
around the time of the binary's merger.
The spin memory, however, grows like $x^{-1/2}$ for compact binaries.
Thus, it has a significant contribution from the early inspiral, as well as
a contribution from around the time of merger.

While it may seem surprising that a nonlinear gravitational effect like the
spin memory would have a large gradual accumulation as the binary inspirals 
from the weak-field regime to the strong-field one, this behavior is similar 
to that of the total angular momentum radiated by a quasicircular compact 
binary.
The Newtonian angular momentum for such a binary also scales with the PN 
parameter $x$ as $x^{-1/2}$.
Because any astrophysical binary will have its components be separated by
a finite distance, its total radiated angular momentum or its spin memory
effect will be finite quantites (i.e., the PN parameter at the time of its
formation will be some nonzero, but small quantity). 

The rate of change of how the spin memory accumulates (what we called the
spin memory mode) turns out to be identical to a 2.5PN-order effect in the
cross polarization of a gravitational waveform for nonspinning compact
binaries.
In addition to giving a concrete interpretation to this unusual nonlinear, but
nonhereditary and nonoscillatory effect, it also provides an avenue by which
the spin memory could be observed in gravitational-wave detectors.
Because of the spin memory's relationship to the extended BMS symmetry, this
would be a promising method by which we can learn, observationally, about the 
appropriate boundary conditions and asymptotic symmetries of isolated 
gravitational systems.

We therefore estimated the signal-to-noise ratio of the spin memory mode 
in advanced LIGO at its design sensitivity and the planned detector, the
Einstein Telescope.
For individual binary mergers, detecting the spin memory mode seems unlikely,
even for the Einstein Telescope.
Instead, if one were to coherently combine measurements of the spin memory
mode from hundreds of gravitational-wave observations of binary-black-hole 
mergers in the Einstein Telescope, the prospects for observing the effect 
look better.
Additional work to refine the strategy for coherently combining measurements
of the spin memory mode will be needed in the future.

Other directions for future work are as follows:
First, it would be of interest to study the detectability of the spin memory
for pulsar timing arrays and space-based gravitational-wave detectors in 
greater detail.
Second, it would be useful to compute higher-order PN corrections to the 
spin memory effect for compact binaries so as to model the effect from the 
last few orbits and merger of compact objects more precisely.
Third, it would also be important to compute the effect in 
numerical relativity simulations of compact binaries and supernovae.

\acknowledgments

I thank Luc Blanchet for asking about the spin memory effect in PN theory,
which prompted this line of inquiry.
I am also grateful to Guillame Faye for his explanations of aspects of the 
PN approach.
I express gratitude to Yanbei Chen for providing insight about how to detect 
the spin memory mode and Leo Stein for discussions about how to measure the 
spin memory and for comments on this paper.
I appreciate correspondence from Zhoujian Cao and Thomas M\"adler on this 
paper.
Finally, I thank Samaya Nissanke for reading this paper and providing valuable 
comments and suggestions.
This work is part of the research program Innovational Research Incentives 
Scheme (Vernieuwingsimpuls), which is financed by the Netherlands Organization
for Scientific Research through the NWO VIDI Grant No.~639.042.612-Nissanke.

\appendix

\section{\label{sec:harmonic_relations} Conventions for pure-spin tensor 
harmonics and relations to spin-weighted spherical harmonics}

The pure-spin vector and tensor harmonics are constructed by applying 
the covariant derivative $D_A$ that is compatible with the 2-sphere metric
$h_{AB}$ to the scalar spherical harmonics $Y_{lm}$ (which are normalized 
with the Condon-Shortley phase convention).
The definitions of the vector harmonics that we use are
\begin{subequations}
\begin{align}
T^{(e),lm}_A = & \frac 1{\sqrt{l(l+1)}}  D_A Y_{lm} \, ,\\
T^{(b),lm}_A = & \frac 1{\sqrt{l(l+1)}}  \epsilon_{AB} D^B Y_{lm} \, ,
\end{align}
\end{subequations}
where we require that $l\geq 1$.
The second-rank tensor harmonics $T^{(e),lm}_{AB}$ and 
$T^{(b),lm}_{AB}$ we will define as
\begin{subequations}
\begin{align}
T^{(e),lm}_{AB} = & \frac 12 \sqrt{\frac{2(l-2)!}{(l+2)!}} 
(2 D_A D_B - h_{AB} D^2) Y_{lm} \, ,\\
T^{(b),lm}_{AB} = & \sqrt{\frac{2(l-2)!}{(l+2)!}}
\epsilon_{C(A} D_{B)} D^C Y_{lm} \, ,
\end{align}
\label{eq:TABharmonics}%
\end{subequations}
which are nonzero for $l\geq 2$.
The highest-rank harmonics we will need are the third-rank tensor harmonics, 
which we define by
\begin{subequations}
\begin{align}
T^{(e),lm}_{ABC} = & 2\sqrt{\frac{(l-3)!}{(l+3)!}} 
(D_A D_B D_C Y_{lm})^{\mathrm{STF}} \, , \\
T^{(b),lm}_{ABC} = & 2\sqrt{\frac{(l-3)!}{(l+3)!}} 
(D_A \epsilon_{DB} D_{C} D^D Y_{lm})^{\mathrm{STF}} \, ,
\end{align}
\end{subequations}
where these harmonics are nonzero if $l\geq 3$.

The pure spin harmonics are complete for smooth vector, second-rank, and 
third-rank symmetric, trace-free tensor fields, and they satisfy the 
orthogonality relationships
\begin{subequations}
\begin{align}
\int d^2\Omega T^{(e),lm}_{A} T_{(e),l'm'}^{A} = & 
\delta_{l'l}\delta_{m'm} \, , \\
\int d^2\Omega T^{(b),lm}_{A} T_{(b),l'm'}^{A} = & 
\delta_{l'l}\delta_{m'm} \, , \\
\int d^2\Omega T^{(e),lm}_{A} T_{(b),l'm'}^{A} = & 0 \, ,
\end{align}
\end{subequations}
for the vector harmonics, 
\begin{subequations}
\begin{align}
\int d^2\Omega T^{(e),lm}_{AB} T_{(e),l'm'}^{AB} = & 
\delta_{l'l}\delta_{m'm} \, , \\
\int d^2\Omega T^{(b),lm}_{AB} T_{(b),l'm'}^{AB} = & 
\delta_{l'l}\delta_{m'm} \, , \\
\int d^2\Omega T^{(e),lm}_{AB} T_{(b),l'm'}^{AB} = & 0 \, ,
\end{align}
\end{subequations}
for the second-rank tensor harmonics, and
\begin{subequations}
\begin{align}
\int d^2\Omega T^{(e),lm}_{ABC} T_{(e),l'm'}^{ABC} = & 
\delta_{l'l}\delta_{m'm} \, , \\
\int d^2\Omega T^{(b),lm}_{ABC} T_{(b),l'm'}^{ABC} = & 
\delta_{l'l}\delta_{m'm} \, , \\
\int d^2\Omega T^{(e),lm}_{ABC} T_{(b),l'm'}^{ABC} = & 0 \, ,
\end{align}
\end{subequations}
for the third-rank tensor harmonics.

Finally, we give relationships between the pure-spin harmonics and the
spin-weighted spherical harmonics, which we define by
\begin{subequations}
\begin{align}
{}_sY_{lm} = & \sqrt{\frac{(l-s)!}{(l+s)!}} \eth^s Y_{lm} & s \geq 0 \, , \\
{}_sY_{lm} = & (-1)^s \sqrt{\frac{(l+s)!}{(l-s)!}} \bar \eth^{-s} Y_{lm} & 
s < 0 \, ,
\end{align}
\end{subequations}
where the operators $\eth$ and $\bar\eth$ act on a function $f$ of spin 
weight $s$ by
\begin{subequations}
\begin{align}
\eth f = -\sin^s\theta(\partial_\theta + i \csc\theta) (f\sin^{-s}\theta) 
\, , \\
\bar \eth f = -\sin^{-s}\theta(\partial_\theta - i \csc\theta) 
(f\sin^{s}\theta) \, .
\end{align}
\end{subequations}
The relationships for the vector harmonics are 
\begin{subequations}
\begin{align}
T^{(e),lm}_{A} = & \frac 1{\sqrt 2} ({}_{-1}Y_{lm}m_A - 
{}_{1}Y_{lm} \bar m_A) \, ,\\
T^{(b),lm}_{A} = & \frac i{\sqrt 2} ({}_{-1}Y_{lm}m_A +
{}_{1}Y_{lm} \bar m_A) \, .
\end{align}
\label{eq:Rank1Spin1Ylm}%
\end{subequations}
For the second-rank tensor harmonics they are 
\begin{subequations}
\begin{align}
T^{(e),lm}_{AB} = & \frac 1{\sqrt 2} ({}_{-2}Y_{lm}m_A m_B + 
{}_{2}Y_{lm} \bar m_A \bar m_B) \, ,\\
T^{(b),lm}_{AB} = & -\frac i{\sqrt 2} ({}_{-2}Y_{lm}m_A m_B -
{}_{2}Y_{lm} \bar m_A \bar m_B) \, ,
\end{align}
\label{eq:Rank2Spin2Ylm}%
\end{subequations}
and for the third-rank tensor harmonics they are
\begin{subequations}
\begin{align}
T^{(e),lm}_{ABC} = & \frac 1{\sqrt 2} ({}_{-3}Y_{lm}m_A m_B m_C - 
{}_{3}Y_{lm} \bar m_A \bar m_B \bar m_C) \, ,\\
T^{(b),lm}_{ABC} = & -\frac i{\sqrt 2} ({}_{-3}Y_{lm}m_A m_B m_C +
{}_{3}Y_{lm} \bar m_A \bar m_B \bar m_C) \, .
\end{align}
\label{eq:Rank3Spin3Ylm}%
\end{subequations}
In the above expressions, the vector $m_A$ has the form 
$m_A \partial^A = \partial_\theta + i\csc\theta \partial_\phi$ in spherical
polar coordinates, and $\bar m_A$ is its complex conjugate (independent of
the coordinate chart).
With these definitions, it can then be easily shown that our convention for
the gravitational-wave polarizations is defined by
\begin{equation}
h = h_+ - i h_\times = C_{AB} \bar m^A \bar m^B \, ,
\end{equation}
where $C_{AB}$ is the strain tensor in the Bondi formalism.

\section{\label{sec:AngMomFlux} Flux of angular momentum}

In this appendix, we compute the flux of angular momentum.
It is given by the $l=1$, $m=0,\pm 1$ moments of the right-hand side of
Eq.~\eqref{eq:NullSpinMemory} when properly normalized and with the 
integral with respect to $u$ removed.
The expression is 
\begin{align}
\frac{dJ_{1,m}}{du} = & \frac 1{64\pi} \int d^2\Omega  
( C^{BC} D_B N_{AC} - N^{BC} D_B C_{AC} \nonumber \\
& + 3N_{AB} D_C C^{BC} - 3C_{AB} D_C N^{BC}) \epsilon^{AD} D_D\bar Y_{1,m} \, .
\end{align}
When written in terms of the radiative multipole moments, the angular momentum 
flux has the form very similar to that shown in Eq.~\eqref{eq:NullSpinMemMom},
but because this expression is specialized to the $l=1$ moments now,
only terms in which $|l'-l''|\leq 1$ and $|m'+m''|\leq 1$ contribute 
to the sum.
As a result there are several significant simplifications that occur.
The most important of which is that the contribution from the set of terms 
begin multiplied by $[1-(-1)^{l'+l''}]$ turn out to be zero and only those 
being multiplied by $[1+(-1)^{l'+l''}]$ contribute.
As a result, the only contributions to the angular momentum flux come from the
moments with $l'=l''$, and the double sum reduces to a single sum over 
multipoles.
After some computation, we find that
\begin{subequations}
\begin{align}
\frac{dJ_{1,0}}{du} = & -\frac{i}{64\pi} \sqrt{\frac 3\pi}
\sum_{l',m'} m' (\bar U_{l'm'} \dot U_{l'm'} + \bar V_{l'm'} \dot V_{l'm'})
\, ,\\
\frac{dJ_{1,\pm 1}}{du} = & \pm \frac{i}{64\pi} 
\sqrt{\frac 3{2\pi}} \sum_{l',m'} \sqrt{(l'\mp m')(l'\pm m'+1)} 
\nonumber \\
& \times (\bar U_{l'm'} \dot U_{l'm'\pm 1} + \bar V_{l'm'} \dot V_{l'm'\pm 1}) 
\, ,
\end{align}
\label{eq:dJ1mdu}%
\end{subequations}
where, as usual, $l'\geq 2$.
Because the flux of angular momentum is real, then the $m=\pm 1$ multipoles 
satisfy the relationship
\begin{equation}
\frac{dJ_{1,1}}{du} = 
- \frac{d\bar J_{1,-1}}{du} \, .
\label{eq:dJduReal}
\end{equation}

Next, we would like to relate these multipole moments to the expression in 
inertial Minkowski coordinates given in~\cite{Thorne1980}. 
The way to do this is described in~\cite{Flanagan:2015pxa}.
First, define the vectors
\begin{equation}
n_i = (\sin\theta\cos\phi,\sin\theta\sin\phi,\cos\theta) \, ,
\qquad e^A_i = D^A n_i \, ,
\end{equation}
and then write the $l=1$, magnetic-parity vector harmonics as 
\begin{equation}
\epsilon^{AB} D_B \bar Y_{1,m} = \omega^{ij}_{1,m} e^A_{[i}n_{j]} 
\end{equation}
for some antisymmetric tensors $\omega_{ij}^{1,m} = \omega_{[ij]}^{1,m}$.
From the expressions for the vector harmonics, the coefficients take the form
\begin{subequations}
\begin{align}
\omega^{xy}_{1,0} = & -\frac 12 \sqrt{\frac 3{\pi}} \, , \qquad
\omega^{xz}_{1,0} = 0 \, , \qquad  \omega^{yz}_{1,0} = 0 \, ,\\
\omega^{xy}_{1,\pm 1} = & 0 \, , \qquad 
\omega^{xz}_{1,\pm 1} = \frac i2 \sqrt{\frac 3{2\pi}} \, , \qquad
\omega^{yz}_{1,\pm 1} = \pm \frac 12 \sqrt{\frac 3{2\pi}} \, .
\end{align}
\label{eq:omegaij}%
\end{subequations}
The multipole moments of the angular momentum flux are related to the 
inertial-Minkowski-coordinate components of the flux by
\begin{equation}
\frac{dJ_{1,m}}{du} = \frac 12 \omega^{ij}_{1,m} 
\frac{dJ_{ij}}{du} \, .
\label{eq:FluxRelation}
\end{equation}
With the relationship
\begin{equation}
\frac{dJ_{i}}{du} = \frac 12 \epsilon_{ijk} 
\frac{dJ_{jk}}{du} \, ,
\end{equation}
one can combine the results of Eqs.~\eqref{eq:dJ1mdu},~\eqref{eq:omegaij},
and~\eqref{eq:FluxRelation} to solve for the inertial Minkowski components
of the flux 
\begin{subequations}
\begin{align}
\frac{dJ_{x}}{du} = & \frac{i}{64\pi} 
\sum_{l',m'} [\sqrt{(l'- m')(l'+m'+1)} 
\nonumber \\
& \times (\bar U_{l'm'} \dot U_{l'm'+1} + \bar V_{l'm'} \dot V_{l'm'+1}) 
\nonumber \\
& + \sqrt{(l'+m')(l'-m'+1)} 
\nonumber \\
& \times (\bar U_{l'm'} \dot U_{l'm'-1} 
+ \bar V_{l'm'} \dot V_{l'm'-1}) ] \, ,\\
\frac{dJ_{y}}{du} = & - \frac{1}{64\pi} 
\sum_{l',m'} [\sqrt{(l'- m')(l'+m'+1)} 
\nonumber \\
& \times (\bar U_{l'm'} \dot U_{l'm'+1} + \bar V_{l'm'} \dot V_{l'm'+1}) 
\nonumber \\
& - \sqrt{(l'+m')(l'-m'+1)} 
\nonumber \\
& \times (\bar U_{l'm'} \dot U_{l'm'-1} 
+ \bar V_{l'm'} \dot V_{l'm'-1}) ] \, ,\\
\frac{dJ_{z}}{du} = & \frac i{32\pi} \sum_{l'm'} m' 
(\bar U_{l'm'} \dot U_{l'm'} + \bar V_{l'm'} \dot V_{l'm'}) \, .
\end{align}
\end{subequations}
These expressions agree with those in Eq.~(4.23) of~\cite{Thorne1980}.
While the individual terms in the sum are not necessarily real, it can be shown
that the total sum is, in fact, real.

\section{\label{sec:SpinMemNR} Comparison with spin memory mode in 
numerical simulations}

\begin{figure}[hbt]
\includegraphics[width=0.99\columnwidth]{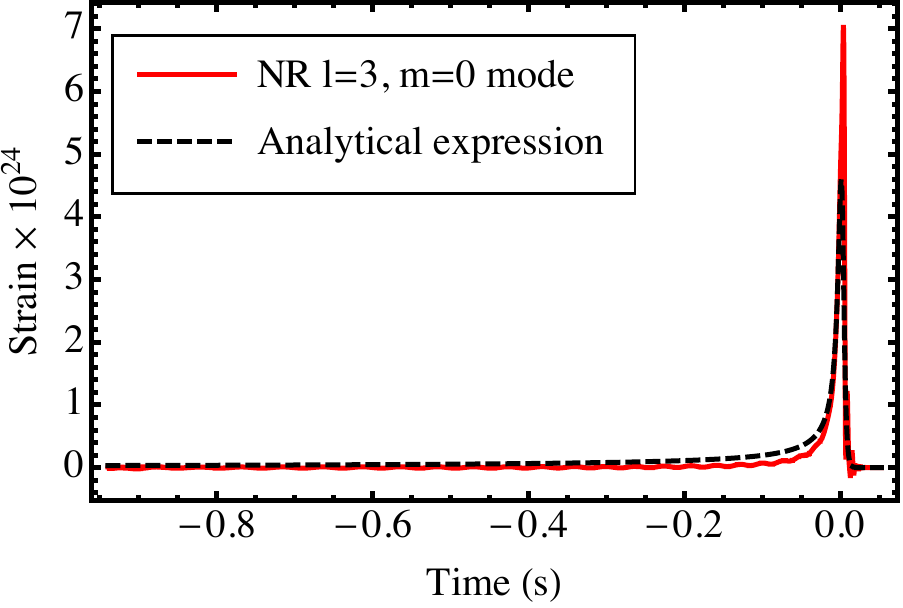}
\caption{\label{fig:SpinMemMode}
Spin memory mode in a numerical relativity simulation (solid red curve)
and that computed from the analytical expression involving the $l=2$,
$m=\pm 2$ modes (black dashed curve).
The waveform is scaled to have parameters like that of the GW150914 event,
which are described in greater detail in the text.}
\end{figure}

\begin{figure}[htb]
\includegraphics[width=0.99\columnwidth]{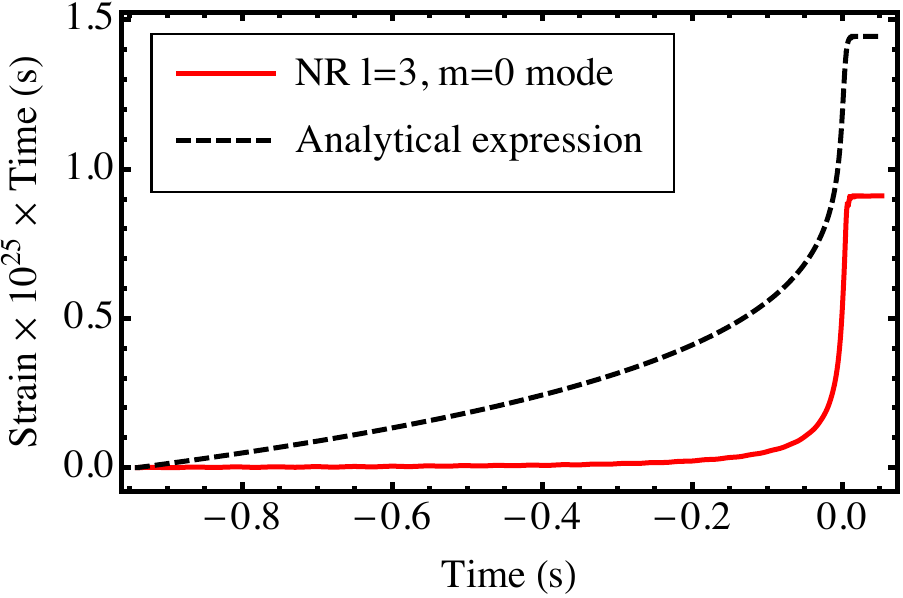}
\caption{\label{fig:SpinMem}
The spin memory effect computed directly from a numerical relativity 
simulation and the analytical expression containing the $l=2$, $m=\pm 2$
by integrating the curves in Fig.~\ref{fig:SpinMemMode} numerically.}
\end{figure}

As a brief exercise in relating analytical and numerical-relativity 
predictions, we compare the spin memory mode from the simulation SXS:BBH:0305 
from the SXS waveform catalog~\cite{SXS:catalog} with the analytical result 
that is given in Eq.~\eqref{eq:SpinMemMode}.
We also compare the $u$ integral of~\eqref{eq:SpinMemMode} and the 
numerical $l=3$, $m=0$ mode.
It is not a strict analytical comparison in the sense that we use the 
$l=2$, $m=\pm 2$ modes from the numerical simulation in our analytical
expression for the spin memory modes; instead, it can be thought of as a
consistency check of the numerical modes with the analytical expectations
for these modes.

One might be concerned that this is a fruitless exercise, because one of the
major caveats noted in~\cite{SXS:catalog} is that the $m=0$ modes are not
captured well by the extrapolation method used to compute the waveforms
at future null infinity.
For the displacement memory mode ($l=2$, $m=0$), the waveform computed
from Eq.~\eqref{eq:DispMemMode} and the $l=2$, $m=0$ mode from the simulation
show large discrepancies between the two (not even qualitative similarities).

Interestingly, the spin memory modes are much more similar.
For illustrative purposes, we rescale the numerical waveforms to have the 
same total mass, luminosity distance, and inclination as GW150914 (as detailed
in Sec.~\ref{sec:detection}).
The result of this comparison is shown in Fig.~\ref{fig:SpinMemMode}.
The solid red curve is the numerical-relativity waveform, and the dashed
black curve is the result calculated using the numerical $l=2$, $m=\pm 2$ 
modes and the analytical expression~\eqref{eq:SpinMemMode}.
The peak of the magnitude of the $l=2$, $m=2$ mode is chosen to be the time
$t=0$ in Figs.~\ref{fig:SpinMemMode} and~\ref{fig:SpinMem}.
The agreement between the two methods is remarkably good, especially
considering how poor it is for the displacement memory mode.
We speculate that it might occur because of the nonhereditary nature of the
spin memory mode or because of the higher PN order of the effect.
It would be interesting to investigate this in greater detail.

To temper any overexuberance about the strength of the agreement, 
we also show the spin memory in Fig.~\ref{fig:SpinMem}, which was computed
by integrating the two curves in Fig.~\ref{fig:SpinMemMode} numerically.
We chose the constants of integration so that both curves went to zero at
the earliest time depicted in the figure.
Even though the peak of the numerical-relativity curve in 
Fig.~\ref{fig:SpinMemMode} is noticeably larger than that of the analytical
curve, the analytical curve has a larger spin memory effect than the 
numerical one does.
This occurs because the numerical curve hovers more closely to zero than
the analytical curve does; as a result, the analytical expression is able
to capture more of the gradual accumulation of the spin memory.
And although the differences between the curves in Figs.~\ref{fig:SpinMemMode}
and~\ref{fig:SpinMem} are larger in the latter than the former, the 
qualitative agreement in both is rather impressive.

\bibliography{Refs}

\begin{thebibliography}{58}%
\makeatletter
\providecommand \@ifxundefined [1]{%
 \@ifx{#1\undefined}
}%
\providecommand \@ifnum [1]{%
 \ifnum #1\expandafter \@firstoftwo
 \else \expandafter \@secondoftwo
 \fi
}%
\providecommand \@ifx [1]{%
 \ifx #1\expandafter \@firstoftwo
 \else \expandafter \@secondoftwo
 \fi
}%
\providecommand \natexlab [1]{#1}%
\providecommand \enquote  [1]{``#1''}%
\providecommand \bibnamefont  [1]{#1}%
\providecommand \bibfnamefont [1]{#1}%
\providecommand \citenamefont [1]{#1}%
\providecommand \href@noop [0]{\@secondoftwo}%
\providecommand \href [0]{\begingroup \@sanitize@url \@href}%
\providecommand \@href[1]{\@@startlink{#1}\@@href}%
\providecommand \@@href[1]{\endgroup#1\@@endlink}%
\providecommand \@sanitize@url [0]{\catcode `\\12\catcode `\$12\catcode
  `\&12\catcode `\#12\catcode `\^12\catcode `\_12\catcode `\%12\relax}%
\providecommand \@@startlink[1]{}%
\providecommand \@@endlink[0]{}%
\providecommand \url  [0]{\begingroup\@sanitize@url \@url }%
\providecommand \@url [1]{\endgroup\@href {#1}{\urlprefix }}%
\providecommand \urlprefix  [0]{URL }%
\providecommand \Eprint [0]{\href }%
\providecommand \doibase [0]{http://dx.doi.org/}%
\providecommand \selectlanguage [0]{\@gobble}%
\providecommand \bibinfo  [0]{\@secondoftwo}%
\providecommand \bibfield  [0]{\@secondoftwo}%
\providecommand \translation [1]{[#1]}%
\providecommand \BibitemOpen [0]{}%
\providecommand \bibitemStop [0]{}%
\providecommand \bibitemNoStop [0]{.\EOS\space}%
\providecommand \EOS [0]{\spacefactor3000\relax}%
\providecommand \BibitemShut  [1]{\csname bibitem#1\endcsname}%
\let\auto@bib@innerbib\@empty
\bibitem [{\citenamefont {{Zel'dovich}}\ and\ \citenamefont
  {{Polnarev}}(1974)}]{Zeldovich1974}%
  \BibitemOpen
  \bibfield  {author} {\bibinfo {author} {\bibfnamefont {Y.~B.}\ \bibnamefont
  {{Zel'dovich}}}\ and\ \bibinfo {author} {\bibfnamefont {A.~G.}\ \bibnamefont
  {{Polnarev}}},\ }\bibfield  {title} {\enquote {\bibinfo {title} {{Radiation
  of gravitational waves by a cluster of superdense stars}},}\ }\href@noop {}
  {\bibfield  {journal} {\bibinfo  {journal} {Sov. Astron.}\ }\textbf {\bibinfo
  {volume} {18}},\ \bibinfo {pages} {17} (\bibinfo {year} {1974})}\BibitemShut
  {NoStop}%
\bibitem [{\citenamefont {{Turner}}(1977)}]{Turner1977}%
  \BibitemOpen
  \bibfield  {author} {\bibinfo {author} {\bibfnamefont {M.~S.}\ \bibnamefont
  {{Turner}}},\ }\bibfield  {title} {\enquote {\bibinfo {title} {{Gravitational
  radiation from point-masses in unbound orbits - Newtonian results}},}\ }\href
  {\doibase 10.1086/155501} {\bibfield  {journal} {\bibinfo  {journal} {\apj}\
  }\textbf {\bibinfo {volume} {216}},\ \bibinfo {pages} {610--619} (\bibinfo
  {year} {1977})}\BibitemShut {NoStop}%
\bibitem [{\citenamefont {{Turner}}\ and\ \citenamefont
  {{Will}}(1978)}]{Turner1978a}%
  \BibitemOpen
  \bibfield  {author} {\bibinfo {author} {\bibfnamefont {M.~S.}\ \bibnamefont
  {{Turner}}}\ and\ \bibinfo {author} {\bibfnamefont {C.~M.}\ \bibnamefont
  {{Will}}},\ }\bibfield  {title} {\enquote {\bibinfo {title} {{Post-Newtonian
  gravitational bremsstrahlung}},}\ }\href {\doibase 10.1086/155996} {\bibfield
   {journal} {\bibinfo  {journal} {\apj}\ }\textbf {\bibinfo {volume} {220}},\
  \bibinfo {pages} {1107--1124} (\bibinfo {year} {1978})}\BibitemShut {NoStop}%
\bibitem [{\citenamefont {{Kovacs}}\ and\ \citenamefont
  {{Thorne}}(1978)}]{Kovacs1978}%
  \BibitemOpen
  \bibfield  {author} {\bibinfo {author} {\bibfnamefont {S.~J.}\ \bibnamefont
  {{Kovacs}}, \bibfnamefont {Jr.}}\ and\ \bibinfo {author} {\bibfnamefont
  {K.~S.}\ \bibnamefont {{Thorne}}},\ }\bibfield  {title} {\enquote {\bibinfo
  {title} {{The generation of gravitational waves. IV - Bremsstrahlung}},}\
  }\href {\doibase 10.1086/156350} {\bibfield  {journal} {\bibinfo  {journal}
  {\apj}\ }\textbf {\bibinfo {volume} {224}},\ \bibinfo {pages} {62--85}
  (\bibinfo {year} {1978})}\BibitemShut {NoStop}%
\bibitem [{\citenamefont {{Turner}}(1978)}]{Turner1978b}%
  \BibitemOpen
  \bibfield  {author} {\bibinfo {author} {\bibfnamefont {M.~S.}\ \bibnamefont
  {{Turner}}},\ }\bibfield  {title} {\enquote {\bibinfo {title} {{Gravitational
  radiation from supernova neutrino bursts}},}\ }\href {\doibase
  10.1038/274565a0} {\bibfield  {journal} {\bibinfo  {journal} {\nat}\ }\textbf
  {\bibinfo {volume} {274}},\ \bibinfo {pages} {565} (\bibinfo {year}
  {1978})}\BibitemShut {NoStop}%
\bibitem [{\citenamefont {Christodoulou}(1991)}]{Christodoulou1991}%
  \BibitemOpen
  \bibfield  {author} {\bibinfo {author} {\bibfnamefont {D.}~\bibnamefont
  {Christodoulou}},\ }\bibfield  {title} {\enquote {\bibinfo {title} {Nonlinear
  nature of gravitation and gravitational-wave experiments},}\ }\href {\doibase
  10.1103/PhysRevLett.67.1486} {\bibfield  {journal} {\bibinfo  {journal}
  {Phys. Rev. Lett.}\ }\textbf {\bibinfo {volume} {67}},\ \bibinfo {pages}
  {1486--1489} (\bibinfo {year} {1991})}\BibitemShut {NoStop}%
\bibitem [{\citenamefont {Wiseman}\ and\ \citenamefont
  {Will}(1991)}]{Wiseman1991}%
  \BibitemOpen
  \bibfield  {author} {\bibinfo {author} {\bibfnamefont {A.~G.}\ \bibnamefont
  {Wiseman}}\ and\ \bibinfo {author} {\bibfnamefont {C.~M.}\ \bibnamefont
  {Will}},\ }\bibfield  {title} {\enquote {\bibinfo {title} {Christodoulou's
  nonlinear gravitational-wave memory: Evaluation in the quadrupole
  approximation},}\ }\href {\doibase 10.1103/PhysRevD.44.R2945} {\bibfield
  {journal} {\bibinfo  {journal} {Phys. Rev. D}\ }\textbf {\bibinfo {volume}
  {44}},\ \bibinfo {pages} {R2945--R2949} (\bibinfo {year} {1991})}\BibitemShut
  {NoStop}%
\bibitem [{\citenamefont {Blanchet}\ and\ \citenamefont
  {Damour}(1992)}]{Blanchet1992}%
  \BibitemOpen
  \bibfield  {author} {\bibinfo {author} {\bibfnamefont {L.}~\bibnamefont
  {Blanchet}}\ and\ \bibinfo {author} {\bibfnamefont {T.}~\bibnamefont
  {Damour}},\ }\bibfield  {title} {\enquote {\bibinfo {title} {Hereditary
  effects in gravitational radiation},}\ }\href {\doibase
  10.1103/PhysRevD.46.4304} {\bibfield  {journal} {\bibinfo  {journal} {Phys.
  Rev. D}\ }\textbf {\bibinfo {volume} {46}},\ \bibinfo {pages} {4304--4319}
  (\bibinfo {year} {1992})}\BibitemShut {NoStop}%
\bibitem [{\citenamefont {Favata}(2009{\natexlab{a}})}]{Favata:2008yd}%
  \BibitemOpen
  \bibfield  {author} {\bibinfo {author} {\bibfnamefont {M.}~\bibnamefont
  {Favata}},\ }\bibfield  {title} {\enquote {\bibinfo {title} {{Post-Newtonian
  corrections to the gravitational-wave memory for quasi-circular, inspiralling
  compact binaries}},}\ }\href {\doibase 10.1103/PhysRevD.80.024002} {\bibfield
   {journal} {\bibinfo  {journal} {Phys. Rev. D}\ }\textbf {\bibinfo {volume}
  {80}},\ \bibinfo {pages} {024002} (\bibinfo {year} {2009}{\natexlab{a}})},\
  \Eprint {http://arxiv.org/abs/0812.0069} {arXiv:0812.0069 [gr-qc]}
  \BibitemShut {NoStop}%
\bibitem [{\citenamefont {Pollney}\ and\ \citenamefont
  {Reisswig}(2011)}]{Pollney2010}%
  \BibitemOpen
  \bibfield  {author} {\bibinfo {author} {\bibfnamefont {D.}~\bibnamefont
  {Pollney}}\ and\ \bibinfo {author} {\bibfnamefont {C.}~\bibnamefont
  {Reisswig}},\ }\bibfield  {title} {\enquote {\bibinfo {title} {{Gravitational
  memory in binary black hole mergers}},}\ }\href {\doibase
  10.1088/2041-8205/732/1/L13} {\bibfield  {journal} {\bibinfo  {journal}
  {Astrophys.~J.~Lett.}\ }\textbf {\bibinfo {volume} {732}},\ \bibinfo {pages}
  {L13} (\bibinfo {year} {2011})},\ \Eprint {http://arxiv.org/abs/1004.4209}
  {arXiv:1004.4209 [gr-qc]} \BibitemShut {NoStop}%
\bibitem [{\citenamefont {Cao}\ and\ \citenamefont {Han}(2016)}]{Cao:2016maz}%
  \BibitemOpen
  \bibfield  {author} {\bibinfo {author} {\bibfnamefont {Z.}~\bibnamefont
  {Cao}}\ and\ \bibinfo {author} {\bibfnamefont {W.}~\bibnamefont {Han}},\
  }\bibfield  {title} {\enquote {\bibinfo {title} {{Inspiral-merger-ringdown
  (2, 0) mode waveforms for aligned-spin black-hole binaries}},}\ }\href
  {\doibase 10.1088/0264-9381/33/15/155011} {\bibfield  {journal} {\bibinfo
  {journal} {Classical Quantum Gravity}\ }\textbf {\bibinfo {volume} {33}},\
  \bibinfo {pages} {155011} (\bibinfo {year} {2016})}\BibitemShut {NoStop}%
\bibitem [{\citenamefont {Braginsky}\ and\ \citenamefont
  {Grishchuk}(1985)}]{Braginsky1985}%
  \BibitemOpen
  \bibfield  {author} {\bibinfo {author} {\bibfnamefont {V.~B.}\ \bibnamefont
  {Braginsky}}\ and\ \bibinfo {author} {\bibfnamefont {L.~P.}\ \bibnamefont
  {Grishchuk}},\ }\bibfield  {title} {\enquote {\bibinfo {title} {Kinematic
  resonance and memory effect in free-mass gravitational antennas},}\
  }\href@noop {} {\bibfield  {journal} {\bibinfo  {journal} {Sov. Phys. JETP}\
  }\textbf {\bibinfo {volume} {89}},\ \bibinfo {pages} {744} (\bibinfo {year}
  {1985})}\BibitemShut {NoStop}%
\bibitem [{\citenamefont {{Braginskii}}\ and\ \citenamefont
  {{Thorne}}(1987)}]{Braginsky1987}%
  \BibitemOpen
  \bibfield  {author} {\bibinfo {author} {\bibfnamefont {V.~B.}\ \bibnamefont
  {{Braginskii}}}\ and\ \bibinfo {author} {\bibfnamefont {K.~S.}\ \bibnamefont
  {{Thorne}}},\ }\bibfield  {title} {\enquote {\bibinfo {title}
  {{Gravitational-wave bursts with memory and experimental prospects}},}\
  }\href {\doibase 10.1038/327123a0} {\bibfield  {journal} {\bibinfo  {journal}
  {\nat}\ }\textbf {\bibinfo {volume} {327}},\ \bibinfo {pages} {123--125}
  (\bibinfo {year} {1987})}\BibitemShut {NoStop}%
\bibitem [{\citenamefont {Wang}\ \emph {et~al.}(2015)\citenamefont {Wang} \emph
  {et~al.}}]{Wang2015}%
  \BibitemOpen
  \bibfield  {author} {\bibinfo {author} {\bibfnamefont {J.~B.}\ \bibnamefont
  {Wang}} \emph {et~al.},\ }\bibfield  {title} {\enquote {\bibinfo {title}
  {Searching for gravitational wave memory bursts with the parkes pulsar timing
  array},}\ }\href {\doibase 10.1093/mnras/stu2137} {\bibfield  {journal}
  {\bibinfo  {journal} {Mon. Not. R. Astron. Soc.}\ }\textbf {\bibinfo {volume}
  {446}},\ \bibinfo {pages} {1657--1671} (\bibinfo {year} {2015})},\ \Eprint
  {http://arxiv.org/abs/1410.3323} {arXiv:1410.3323 [astro-ph.GA]} \BibitemShut
  {NoStop}%
\bibitem [{\citenamefont {Arzoumanian}\ \emph {et~al.}(2015)\citenamefont
  {Arzoumanian} \emph {et~al.}}]{Arzoumanian2015}%
  \BibitemOpen
  \bibfield  {author} {\bibinfo {author} {\bibfnamefont {Z.}~\bibnamefont
  {Arzoumanian}} \emph {et~al.} (\bibinfo {collaboration} {NANOGrav}),\
  }\bibfield  {title} {\enquote {\bibinfo {title} {{NANOGrav Constraints on
  Gravitational Wave Bursts with Memory}},}\ }\href {\doibase
  10.1088/0004-637X/810/2/150} {\bibfield  {journal} {\bibinfo  {journal}
  {Astrophys. J.}\ }\textbf {\bibinfo {volume} {810}},\ \bibinfo {pages} {150}
  (\bibinfo {year} {2015})},\ \Eprint {http://arxiv.org/abs/1501.05343}
  {arXiv:1501.05343 [astro-ph.GA]} \BibitemShut {NoStop}%
\bibitem [{\citenamefont {Abbott}\ \emph {et~al.}(2009)\citenamefont {Abbott}
  \emph {et~al.}}]{Abbott:2007kv}%
  \BibitemOpen
  \bibfield  {author} {\bibinfo {author} {\bibfnamefont {B.~P.}\ \bibnamefont
  {Abbott}} \emph {et~al.} (\bibinfo {collaboration} {LIGO Scientific}),\
  }\bibfield  {title} {\enquote {\bibinfo {title} {{LIGO: The Laser
  interferometer gravitational-wave observatory}},}\ }\href {\doibase
  10.1088/0034-4885/72/7/076901} {\bibfield  {journal} {\bibinfo  {journal}
  {Rep. Prog. Phys.}\ }\textbf {\bibinfo {volume} {72}},\ \bibinfo {pages}
  {076901} (\bibinfo {year} {2009})},\ \Eprint {http://arxiv.org/abs/0711.3041}
  {arXiv:0711.3041 [gr-qc]} \BibitemShut {NoStop}%
\bibitem [{\citenamefont {Acernese}\ \emph {et~al.}(2015)\citenamefont
  {Acernese} \emph {et~al.}}]{TheVirgo:2014hva}%
  \BibitemOpen
  \bibfield  {author} {\bibinfo {author} {\bibfnamefont {F.}~\bibnamefont
  {Acernese}} \emph {et~al.} (\bibinfo {collaboration} {VIRGO}),\ }\bibfield
  {title} {\enquote {\bibinfo {title} {{Advanced Virgo: a second-generation
  interferometric gravitational wave detector}},}\ }\href {\doibase
  10.1088/0264-9381/32/2/024001} {\bibfield  {journal} {\bibinfo  {journal}
  {Classical Quantum Gravity}\ }\textbf {\bibinfo {volume} {32}},\ \bibinfo
  {pages} {024001} (\bibinfo {year} {2015})},\ \Eprint
  {http://arxiv.org/abs/1408.3978} {arXiv:1408.3978 [gr-qc]} \BibitemShut
  {NoStop}%
\bibitem [{\citenamefont {Aso}\ \emph {et~al.}(2013)\citenamefont {Aso},
  \citenamefont {Michimura}, \citenamefont {Somiya}, \citenamefont {Ando},
  \citenamefont {Miyakawa}, \citenamefont {Sekiguchi}, \citenamefont
  {Tatsumi},\ and\ \citenamefont {Yamamoto}}]{Aso:2013eba}%
  \BibitemOpen
  \bibfield  {author} {\bibinfo {author} {\bibfnamefont {Y.}~\bibnamefont
  {Aso}}, \bibinfo {author} {\bibfnamefont {Y.}~\bibnamefont {Michimura}},
  \bibinfo {author} {\bibfnamefont {K.}~\bibnamefont {Somiya}}, \bibinfo
  {author} {\bibfnamefont {M.}~\bibnamefont {Ando}}, \bibinfo {author}
  {\bibfnamefont {O.}~\bibnamefont {Miyakawa}}, \bibinfo {author}
  {\bibfnamefont {T.}~\bibnamefont {Sekiguchi}}, \bibinfo {author}
  {\bibfnamefont {D.}~\bibnamefont {Tatsumi}}, \ and\ \bibinfo {author}
  {\bibfnamefont {H.}~\bibnamefont {Yamamoto}} (\bibinfo {collaboration}
  {KAGRA}),\ }\bibfield  {title} {\enquote {\bibinfo {title} {{Interferometer
  design of the KAGRA gravitational wave detector}},}\ }\href {\doibase
  10.1103/PhysRevD.88.043007} {\bibfield  {journal} {\bibinfo  {journal} {Phys.
  Rev. D}\ }\textbf {\bibinfo {volume} {88}},\ \bibinfo {pages} {043007}
  (\bibinfo {year} {2013})},\ \Eprint {http://arxiv.org/abs/1306.6747}
  {arXiv:1306.6747 [gr-qc]} \BibitemShut {NoStop}%
\bibitem [{\citenamefont {Lasky}\ \emph {et~al.}(2016)\citenamefont {Lasky},
  \citenamefont {Thrane}, \citenamefont {Levin}, \citenamefont {Blackman},\
  and\ \citenamefont {Chen}}]{Lasky:2016knh}%
  \BibitemOpen
  \bibfield  {author} {\bibinfo {author} {\bibfnamefont {P.~D.}\ \bibnamefont
  {Lasky}}, \bibinfo {author} {\bibfnamefont {E.}~\bibnamefont {Thrane}},
  \bibinfo {author} {\bibfnamefont {Y.}~\bibnamefont {Levin}}, \bibinfo
  {author} {\bibfnamefont {J.}~\bibnamefont {Blackman}}, \ and\ \bibinfo
  {author} {\bibfnamefont {Y.}~\bibnamefont {Chen}},\ }\bibfield  {title}
  {\enquote {\bibinfo {title} {{Detecting gravitational-wave memory with LIGO:
  implications of GW150914}},}\ }\href {\doibase
  10.1103/PhysRevLett.117.061102} {\bibfield  {journal} {\bibinfo  {journal}
  {Phys. Rev. Lett.}\ }\textbf {\bibinfo {volume} {117}},\ \bibinfo {pages}
  {061102} (\bibinfo {year} {2016})},\ \Eprint
  {http://arxiv.org/abs/1605.01415} {arXiv:1605.01415 [astro-ph.HE]}
  \BibitemShut {NoStop}%
\bibitem [{\citenamefont {Audley}\ \emph {et~al.}(2017)\citenamefont {Audley}
  \emph {et~al.}}]{Audley:2017drz}%
  \BibitemOpen
  \bibfield  {author} {\bibinfo {author} {\bibfnamefont {H.}~\bibnamefont
  {Audley}} \emph {et~al.},\ }\bibfield  {title} {\enquote {\bibinfo {title}
  {{Laser Interferometer Space Antenna}},}\ }\href@noop {} {\  (\bibinfo {year}
  {2017})},\ \Eprint {http://arxiv.org/abs/1702.00786} {arXiv:1702.00786
  [astro-ph.IM]} \BibitemShut {NoStop}%
\bibitem [{\citenamefont {Favata}(2009{\natexlab{b}})}]{Favata:2009ii}%
  \BibitemOpen
  \bibfield  {author} {\bibinfo {author} {\bibfnamefont {M.}~\bibnamefont
  {Favata}},\ }\bibfield  {title} {\enquote {\bibinfo {title} {{Nonlinear
  gravitational-wave memory from binary black hole mergers}},}\ }\href
  {\doibase 10.1088/0004-637X/696/2/L159} {\bibfield  {journal} {\bibinfo
  {journal} {Astrophys. J.}\ }\textbf {\bibinfo {volume} {696}},\ \bibinfo
  {pages} {L159--L162} (\bibinfo {year} {2009}{\natexlab{b}})},\ \Eprint
  {http://arxiv.org/abs/0902.3660} {arXiv:0902.3660 [astro-ph.SR]} \BibitemShut
  {NoStop}%
\bibitem [{\citenamefont {{Bondi}}\ \emph {et~al.}(1962)\citenamefont
  {{Bondi}}, \citenamefont {{van der Burg}},\ and\ \citenamefont
  {{Metzner}}}]{Bondi1962}%
  \BibitemOpen
  \bibfield  {author} {\bibinfo {author} {\bibfnamefont {H.}~\bibnamefont
  {{Bondi}}}, \bibinfo {author} {\bibfnamefont {M.~G.~J.}\ \bibnamefont {{van
  der Burg}}}, \ and\ \bibinfo {author} {\bibfnamefont {A.~W.~K.}\ \bibnamefont
  {{Metzner}}},\ }\bibfield  {title} {\enquote {\bibinfo {title}
  {{Gravitational Waves in General Relativity. VII. Waves from Axi-Symmetric
  Isolated Systems}},}\ }\href {\doibase 10.1098/rspa.1962.0161} {\bibfield
  {journal} {\bibinfo  {journal} {Proc. R. Soc. Lond. A}\ }\textbf {\bibinfo
  {volume} {269}},\ \bibinfo {pages} {21--52} (\bibinfo {year}
  {1962})}\BibitemShut {NoStop}%
\bibitem [{\citenamefont {{Sachs}}(1962)}]{Sachs1962a}%
  \BibitemOpen
  \bibfield  {author} {\bibinfo {author} {\bibfnamefont {R.~K.}\ \bibnamefont
  {{Sachs}}},\ }\bibfield  {title} {\enquote {\bibinfo {title} {{Gravitational
  Waves in General Relativity. VIII. Waves in Asymptotically Flat
  Space-Time}},}\ }\href {\doibase 10.1098/rspa.1962.0206} {\bibfield
  {journal} {\bibinfo  {journal} {Proc. R. Soc. Lond. A}\ }\textbf {\bibinfo
  {volume} {270}},\ \bibinfo {pages} {103--126} (\bibinfo {year}
  {1962})}\BibitemShut {NoStop}%
\bibitem [{\citenamefont {Sachs}(1962)}]{Sachs1962b}%
  \BibitemOpen
  \bibfield  {author} {\bibinfo {author} {\bibfnamefont {R.}~\bibnamefont
  {Sachs}},\ }\bibfield  {title} {\enquote {\bibinfo {title} {Asymptotic
  symmetries in gravitational theory},}\ }\href {\doibase
  10.1103/PhysRev.128.2851} {\bibfield  {journal} {\bibinfo  {journal} {Phys.
  Rev.}\ }\textbf {\bibinfo {volume} {128}},\ \bibinfo {pages} {2851--2864}
  (\bibinfo {year} {1962})}\BibitemShut {NoStop}%
\bibitem [{\citenamefont {Strominger}\ and\ \citenamefont
  {Zhiboedov}(2016{\natexlab{a}})}]{Strominger:2014pwa}%
  \BibitemOpen
  \bibfield  {author} {\bibinfo {author} {\bibfnamefont {A.}~\bibnamefont
  {Strominger}}\ and\ \bibinfo {author} {\bibfnamefont {A.}~\bibnamefont
  {Zhiboedov}},\ }\bibfield  {title} {\enquote {\bibinfo {title}
  {{Gravitational Memory, BMS Supertranslations and Soft Theorems}},}\ }\href
  {\doibase 10.1007/JHEP01(2016)086} {\bibfield  {journal} {\bibinfo  {journal}
  {J. High Energy Phys.}\ }\textbf {\bibinfo {volume} {01}},\ \bibinfo {pages}
  {086} (\bibinfo {year} {2016}{\natexlab{a}})},\ \Eprint
  {http://arxiv.org/abs/1411.5745} {arXiv:1411.5745 [hep-th]} \BibitemShut
  {NoStop}%
\bibitem [{\citenamefont {Flanagan}\ and\ \citenamefont
  {Nichols}(2017)}]{Flanagan:2015pxa}%
  \BibitemOpen
  \bibfield  {author} {\bibinfo {author} {\bibfnamefont {{\'E}.~{\'E}.}\
  \bibnamefont {Flanagan}}\ and\ \bibinfo {author} {\bibfnamefont {D.~A.}\
  \bibnamefont {Nichols}},\ }\bibfield  {title} {\enquote {\bibinfo {title}
  {{Conserved charges of the extended Bondi-Metzner-Sachs algebra}},}\ }\href
  {\doibase 10.1103/PhysRevD.95.044002} {\bibfield  {journal} {\bibinfo
  {journal} {Phys. Rev. D}\ }\textbf {\bibinfo {volume} {95}},\ \bibinfo
  {pages} {044002} (\bibinfo {year} {2017})},\ \Eprint
  {http://arxiv.org/abs/1510.03386} {arXiv:1510.03386 [hep-th]} \BibitemShut
  {NoStop}%
\bibitem [{\citenamefont {Ashtekar}(2014)}]{Ashtekar:2014zsa}%
  \BibitemOpen
  \bibfield  {author} {\bibinfo {author} {\bibfnamefont {A.}~\bibnamefont
  {Ashtekar}},\ }\bibfield  {title} {\enquote {\bibinfo {title} {{Geometry and
  Physics of Null Infinity}},}\ }\href@noop {} {\  (\bibinfo {year} {2014})},\
  \Eprint {http://arxiv.org/abs/1409.1800} {arXiv:1409.1800 [gr-qc]}
  \BibitemShut {NoStop}%
\bibitem [{\citenamefont {Barnich}\ and\ \citenamefont
  {Troessaert}(2010{\natexlab{a}})}]{Barnich2009}%
  \BibitemOpen
  \bibfield  {author} {\bibinfo {author} {\bibfnamefont {G.}~\bibnamefont
  {Barnich}}\ and\ \bibinfo {author} {\bibfnamefont {C.}~\bibnamefont
  {Troessaert}},\ }\bibfield  {title} {\enquote {\bibinfo {title} {{Symmetries
  of asymptotically flat 4 dimensional spacetimes at null infinity
  revisited}},}\ }\href {\doibase 10.1103/PhysRevLett.105.111103} {\bibfield
  {journal} {\bibinfo  {journal} {Phys. Rev. Lett.}\ }\textbf {\bibinfo
  {volume} {105}},\ \bibinfo {pages} {111103} (\bibinfo {year}
  {2010}{\natexlab{a}})},\ \Eprint {http://arxiv.org/abs/0909.2617}
  {arXiv:0909.2617 [gr-qc]} \BibitemShut {NoStop}%
\bibitem [{\citenamefont {Barnich}\ and\ \citenamefont
  {Troessaert}(2010{\natexlab{b}})}]{Barnich2010}%
  \BibitemOpen
  \bibfield  {author} {\bibinfo {author} {\bibfnamefont {G.}~\bibnamefont
  {Barnich}}\ and\ \bibinfo {author} {\bibfnamefont {C.}~\bibnamefont
  {Troessaert}},\ }\bibfield  {title} {\enquote {\bibinfo {title} {{Aspects of
  the BMS/CFT correspondence}},}\ }\href {\doibase 10.1007/JHEP05(2010)062}
  {\bibfield  {journal} {\bibinfo  {journal} {J. High Energy Phys.}\ }\textbf
  {\bibinfo {volume} {05}},\ \bibinfo {pages} {062} (\bibinfo {year}
  {2010}{\natexlab{b}})},\ \Eprint {http://arxiv.org/abs/1001.1541}
  {arXiv:1001.1541 [hep-th]} \BibitemShut {NoStop}%
\bibitem [{\citenamefont {Banks}(2003)}]{Banks2003}%
  \BibitemOpen
  \bibfield  {author} {\bibinfo {author} {\bibfnamefont {T.}~\bibnamefont
  {Banks}},\ }\bibfield  {title} {\enquote {\bibinfo {title} {{A Critique of
  pure string theory: Heterodox opinions of diverse dimensions}},}\ }\href@noop
  {} {\  (\bibinfo {year} {2003})},\ \bibinfo {note} {(see footnote 17)},\
  \Eprint {http://arxiv.org/abs/hep-th/0306074} {arXiv:hep-th/0306074 [hep-th]}
  \BibitemShut {NoStop}%
\bibitem [{\citenamefont {Pasterski}\ \emph {et~al.}(2015)\citenamefont
  {Pasterski}, \citenamefont {Strominger},\ and\ \citenamefont
  {Zhiboedov}}]{Pasterski:2015tva}%
  \BibitemOpen
  \bibfield  {author} {\bibinfo {author} {\bibfnamefont {S.}~\bibnamefont
  {Pasterski}}, \bibinfo {author} {\bibfnamefont {A.}~\bibnamefont
  {Strominger}}, \ and\ \bibinfo {author} {\bibfnamefont {A.}~\bibnamefont
  {Zhiboedov}},\ }\bibfield  {title} {\enquote {\bibinfo {title} {{New
  Gravitational Memories}},}\ }\href@noop {} {\  (\bibinfo {year} {2015})},\
  \Eprint {http://arxiv.org/abs/1502.06120} {arXiv:1502.06120 [hep-th]}
  \BibitemShut {NoStop}%
\bibitem [{\citenamefont {Strominger}\ and\ \citenamefont
  {Zhiboedov}(2016{\natexlab{b}})}]{Strominger:2016wns}%
  \BibitemOpen
  \bibfield  {author} {\bibinfo {author} {\bibfnamefont {A.}~\bibnamefont
  {Strominger}}\ and\ \bibinfo {author} {\bibfnamefont {A.}~\bibnamefont
  {Zhiboedov}},\ }\bibfield  {title} {\enquote {\bibinfo {title}
  {{Superrotations and Black Hole Pair Creation}},}\ }\href@noop {} {\
  (\bibinfo {year} {2016}{\natexlab{b}})},\ \Eprint
  {http://arxiv.org/abs/1610.00639} {arXiv:1610.00639 [hep-th]} \BibitemShut
  {NoStop}%
\bibitem [{\citenamefont {Flanagan}\ \emph {et~al.}(2017)\citenamefont
  {Flanagan}, \citenamefont {Harte},\ and\ \citenamefont
  {Nichols}}]{FlanaganInPrep}%
  \BibitemOpen
  \bibfield  {author} {\bibinfo {author} {\bibfnamefont {{\'E}.~{\'E}.}\
  \bibnamefont {Flanagan}}, \bibinfo {author} {\bibfnamefont {A.~I.}\
  \bibnamefont {Harte}}, \ and\ \bibinfo {author} {\bibfnamefont {D.~A.}\
  \bibnamefont {Nichols}},\ }\href@noop {} {\enquote {\bibinfo {title}
  {{Gravitational-wave memory observables}},}\ } (\bibinfo {year} {2017}),\
  \bibinfo {note} {{In Preparation}}\BibitemShut {NoStop}%
\bibitem [{\citenamefont {Abbott}\ \emph
  {et~al.}(2016{\natexlab{a}})\citenamefont {Abbott} \emph
  {et~al.}}]{TheLIGOScientific:2016pea}%
  \BibitemOpen
  \bibfield  {author} {\bibinfo {author} {\bibfnamefont {B.~P.}\ \bibnamefont
  {Abbott}} \emph {et~al.} (\bibinfo {collaboration} {Virgo, LIGO
  Scientific}),\ }\bibfield  {title} {\enquote {\bibinfo {title} {{Binary Black
  Hole Mergers in the first Advanced LIGO Observing Run}},}\ }\href {\doibase
  10.1103/PhysRevX.6.041015} {\bibfield  {journal} {\bibinfo  {journal} {Phys.
  Rev. X}\ }\textbf {\bibinfo {volume} {6}},\ \bibinfo {pages} {041015}
  (\bibinfo {year} {2016}{\natexlab{a}})},\ \Eprint
  {http://arxiv.org/abs/1606.04856} {arXiv:1606.04856 [gr-qc]} \BibitemShut
  {NoStop}%
\bibitem [{\citenamefont {Bishop}\ \emph {et~al.}(1996)\citenamefont {Bishop},
  \citenamefont {Gomez}, \citenamefont {Lehner},\ and\ \citenamefont
  {Winicour}}]{Bishop:1996gt}%
  \BibitemOpen
  \bibfield  {author} {\bibinfo {author} {\bibfnamefont {N.~T.}\ \bibnamefont
  {Bishop}}, \bibinfo {author} {\bibfnamefont {R.}~\bibnamefont {Gomez}},
  \bibinfo {author} {\bibfnamefont {L.}~\bibnamefont {Lehner}}, \ and\ \bibinfo
  {author} {\bibfnamefont {J.}~\bibnamefont {Winicour}},\ }\bibfield  {title}
  {\enquote {\bibinfo {title} {{Cauchy characteristic extraction in numerical
  relativity}},}\ }\href {\doibase 10.1103/PhysRevD.54.6153} {\bibfield
  {journal} {\bibinfo  {journal} {Phys. Rev. D}\ }\textbf {\bibinfo {volume}
  {54}},\ \bibinfo {pages} {6153--6165} (\bibinfo {year} {1996})}\BibitemShut
  {NoStop}%
\bibitem [{\citenamefont {Reisswig}\ \emph {et~al.}(2009)\citenamefont
  {Reisswig}, \citenamefont {Bishop}, \citenamefont {Pollney},\ and\
  \citenamefont {Szilagyi}}]{Reisswig:2009us}%
  \BibitemOpen
  \bibfield  {author} {\bibinfo {author} {\bibfnamefont {C.}~\bibnamefont
  {Reisswig}}, \bibinfo {author} {\bibfnamefont {N.~T.}\ \bibnamefont
  {Bishop}}, \bibinfo {author} {\bibfnamefont {D.}~\bibnamefont {Pollney}}, \
  and\ \bibinfo {author} {\bibfnamefont {B.}~\bibnamefont {Szilagyi}},\
  }\bibfield  {title} {\enquote {\bibinfo {title} {{Unambiguous determination
  of gravitational waveforms from binary black hole mergers}},}\ }\href
  {\doibase 10.1103/PhysRevLett.103.221101} {\bibfield  {journal} {\bibinfo
  {journal} {Phys. Rev. Lett.}\ }\textbf {\bibinfo {volume} {103}},\ \bibinfo
  {pages} {221101} (\bibinfo {year} {2009})},\ \Eprint
  {http://arxiv.org/abs/0907.2637} {arXiv:0907.2637 [gr-qc]} \BibitemShut
  {NoStop}%
\bibitem [{\citenamefont {Handmer}\ \emph {et~al.}(2016)\citenamefont
  {Handmer}, \citenamefont {Szil{\'a}gyi},\ and\ \citenamefont
  {Winicour}}]{Handmer:2016mls}%
  \BibitemOpen
  \bibfield  {author} {\bibinfo {author} {\bibfnamefont {C.~J.}\ \bibnamefont
  {Handmer}}, \bibinfo {author} {\bibfnamefont {B.}~\bibnamefont
  {Szil{\'a}gyi}}, \ and\ \bibinfo {author} {\bibfnamefont {J.}~\bibnamefont
  {Winicour}},\ }\bibfield  {title} {\enquote {\bibinfo {title} {{Spectral
  Cauchy Characteristic Extraction of strain, news and gravitational radiation
  flux}},}\ }\href {\doibase 10.1088/0264-9381/33/22/225007} {\bibfield
  {journal} {\bibinfo  {journal} {Classical Quantum Gravity}\ }\textbf
  {\bibinfo {volume} {33}},\ \bibinfo {pages} {225007} (\bibinfo {year}
  {2016})},\ \Eprint {http://arxiv.org/abs/1605.04332} {arXiv:1605.04332
  [gr-qc]} \BibitemShut {NoStop}%
\bibitem [{\citenamefont {Bieri}\ and\ \citenamefont
  {Garfinkle}(2014)}]{Bieri:2013ada}%
  \BibitemOpen
  \bibfield  {author} {\bibinfo {author} {\bibfnamefont {L.}~\bibnamefont
  {Bieri}}\ and\ \bibinfo {author} {\bibfnamefont {D.}~\bibnamefont
  {Garfinkle}},\ }\bibfield  {title} {\enquote {\bibinfo {title} {{Perturbative
  and gauge invariant treatment of gravitational wave memory}},}\ }\href
  {\doibase 10.1103/PhysRevD.89.084039} {\bibfield  {journal} {\bibinfo
  {journal} {Phys. Rev. D}\ }\textbf {\bibinfo {volume} {89}},\ \bibinfo
  {pages} {084039} (\bibinfo {year} {2014})},\ \Eprint
  {http://arxiv.org/abs/1312.6871} {arXiv:1312.6871 [gr-qc]} \BibitemShut
  {NoStop}%
\bibitem [{\citenamefont {Blanchet}(2014)}]{Blanchet:2006zz}%
  \BibitemOpen
  \bibfield  {author} {\bibinfo {author} {\bibfnamefont {L.}~\bibnamefont
  {Blanchet}},\ }\bibfield  {title} {\enquote {\bibinfo {title} {{Gravitational
  radiation from post-Newtonian sources and inspiralling compact binaries}},}\
  }\href@noop {} {\bibfield  {journal} {\bibinfo  {journal} {Living Rev.
  Relativ.}\ }\textbf {\bibinfo {volume} {17}},\ \bibinfo {pages} {2} (\bibinfo
  {year} {2014})}\BibitemShut {NoStop}%
\bibitem [{\citenamefont {Arun}\ \emph {et~al.}(2004)\citenamefont {Arun},
  \citenamefont {Blanchet}, \citenamefont {Iyer},\ and\ \citenamefont
  {Qusailah}}]{Arun:2004ff}%
  \BibitemOpen
  \bibfield  {author} {\bibinfo {author} {\bibfnamefont {K.~G.}\ \bibnamefont
  {Arun}}, \bibinfo {author} {\bibfnamefont {L.}~\bibnamefont {Blanchet}},
  \bibinfo {author} {\bibfnamefont {B.~R.}\ \bibnamefont {Iyer}}, \ and\
  \bibinfo {author} {\bibfnamefont {M.~S.~S.}\ \bibnamefont {Qusailah}},\
  }\bibfield  {title} {\enquote {\bibinfo {title} {{The 2.5PN gravitational
  wave polarisations from inspiralling compact binaries in circular orbits}},}\
  }\href {\doibase 10.1088/0264-9381/21/15/010} {\bibfield  {journal} {\bibinfo
   {journal} {Classical Quantum Gravity}\ }\textbf {\bibinfo {volume} {21}},\
  \bibinfo {pages} {3771--3802} (\bibinfo {year} {2004})},\ \bibinfo {note}
  {[Erratum: ibid, 22, 3115 (2005)]},\ \Eprint
  {http://arxiv.org/abs/gr-qc/0404085} {arXiv:gr-qc/0404085 [gr-qc]}
  \BibitemShut {NoStop}%
\bibitem [{SXS()}]{SXS:catalog}%
  \BibitemOpen
  \href@noop {} {}\bibinfo {howpublished}
  {\url{http://www.black-holes.org/waveforms}}\BibitemShut {NoStop}%
\bibitem [{\citenamefont {Punturo}\ \emph {et~al.}(2010)\citenamefont {Punturo}
  \emph {et~al.}}]{Punturo:2010zz}%
  \BibitemOpen
  \bibfield  {author} {\bibinfo {author} {\bibfnamefont {M.}~\bibnamefont
  {Punturo}} \emph {et~al.},\ }\bibfield  {title} {\enquote {\bibinfo {title}
  {{The Einstein Telescope: A third-generation gravitational wave
  observatory}},}\ }\bibfield  {booktitle} {\emph {\bibinfo {booktitle}
  {{Proceedings, 14th Workshop on Gravitational wave data analysis (GWDAW-14):
  Rome, Italy, January 26-29, 2010}}},\ }\href {\doibase
  10.1088/0264-9381/27/19/194002} {\bibfield  {journal} {\bibinfo  {journal}
  {Classical Quantum Gravity}\ }\textbf {\bibinfo {volume} {27}},\ \bibinfo
  {pages} {194002} (\bibinfo {year} {2010})}\BibitemShut {NoStop}%
\bibitem [{\citenamefont {Thorne}(1980)}]{Thorne1980}%
  \BibitemOpen
  \bibfield  {author} {\bibinfo {author} {\bibfnamefont {K.~S.}\ \bibnamefont
  {Thorne}},\ }\bibfield  {title} {\enquote {\bibinfo {title} {Multipole
  expansions of gravitational radiation},}\ }\href {\doibase
  10.1103/RevModPhys.52.299} {\bibfield  {journal} {\bibinfo  {journal} {Rev.
  Mod. Phys.}\ }\textbf {\bibinfo {volume} {52}},\ \bibinfo {pages} {299--339}
  (\bibinfo {year} {1980})}\BibitemShut {NoStop}%
\bibitem [{\citenamefont {Wald}(1984)}]{Wald:1984rg}%
  \BibitemOpen
  \bibfield  {author} {\bibinfo {author} {\bibfnamefont {R.~M.}\ \bibnamefont
  {Wald}},\ }\href {\doibase 10.7208/chicago/9780226870373.001.0001} {\emph
  {\bibinfo {title} {{General Relativity}}}}\ (\bibinfo  {publisher} {{Chicago
  University Press}},\ \bibinfo {address} {{Chicago, USA}},\ \bibinfo {year}
  {1984})\BibitemShut {NoStop}%
\bibitem [{\citenamefont {Stewart}(1994)}]{Stewart:1990uf}%
  \BibitemOpen
  \bibfield  {author} {\bibinfo {author} {\bibfnamefont {J.~M.}\ \bibnamefont
  {Stewart}},\ }\href@noop {} {\emph {\bibinfo {title} {{Advanced general
  relativity}}}}\ (\bibinfo  {publisher} {Cambridge University Press},\
  \bibinfo {address} {{Cambridge, U.K.}},\ \bibinfo {year} {1994})\BibitemShut
  {NoStop}%
\bibitem [{\citenamefont {M{\"a}dler}\ and\ \citenamefont
  {Winicour}(2016{\natexlab{a}})}]{Madler:2016xju}%
  \BibitemOpen
  \bibfield  {author} {\bibinfo {author} {\bibfnamefont {T.}~\bibnamefont
  {M{\"a}dler}}\ and\ \bibinfo {author} {\bibfnamefont {J.}~\bibnamefont
  {Winicour}},\ }\bibfield  {title} {\enquote {\bibinfo {title} {{Bondi-Sachs
  Formalism}},}\ }\href {\doibase 10.4249/scholarpedia.33528} {\bibfield
  {journal} {\bibinfo  {journal} {Scholarpedia}\ }\textbf {\bibinfo {volume}
  {11}},\ \bibinfo {pages} {33528} (\bibinfo {year} {2016}{\natexlab{a}})},\
  \Eprint {http://arxiv.org/abs/1609.01731} {arXiv:1609.01731 [gr-qc]}
  \BibitemShut {NoStop}%
\bibitem [{\citenamefont {Flanagan}\ and\ \citenamefont
  {Hughes}(2005)}]{Flanagan:2005yc}%
  \BibitemOpen
  \bibfield  {author} {\bibinfo {author} {\bibfnamefont {E.~E.}\ \bibnamefont
  {Flanagan}}\ and\ \bibinfo {author} {\bibfnamefont {S.~A.}\ \bibnamefont
  {Hughes}},\ }\bibfield  {title} {\enquote {\bibinfo {title} {{The Basics of
  gravitational wave theory}},}\ }\href {\doibase 10.1088/1367-2630/7/1/204}
  {\bibfield  {journal} {\bibinfo  {journal} {New J. Phys.}\ }\textbf {\bibinfo
  {volume} {7}},\ \bibinfo {pages} {204} (\bibinfo {year} {2005})},\ \Eprint
  {http://arxiv.org/abs/gr-qc/0501041} {arXiv:gr-qc/0501041 [gr-qc]}
  \BibitemShut {NoStop}%
\bibitem [{\citenamefont {M{\"a}dler}\ and\ \citenamefont
  {Winicour}(2016{\natexlab{b}})}]{Madler:2016ggp}%
  \BibitemOpen
  \bibfield  {author} {\bibinfo {author} {\bibfnamefont {T.}~\bibnamefont
  {M{\"a}dler}}\ and\ \bibinfo {author} {\bibfnamefont {J.}~\bibnamefont
  {Winicour}},\ }\bibfield  {title} {\enquote {\bibinfo {title} {{The sky
  pattern of the linearized gravitational memory effect}},}\ }\href {\doibase
  10.1088/0264-9381/33/17/175006} {\bibfield  {journal} {\bibinfo  {journal}
  {Classical Quantum Gravity}\ }\textbf {\bibinfo {volume} {33}},\ \bibinfo
  {pages} {175006} (\bibinfo {year} {2016}{\natexlab{b}})},\ \Eprint
  {http://arxiv.org/abs/1605.01273} {arXiv:1605.01273 [gr-qc]} \BibitemShut
  {NoStop}%
\bibitem [{\citenamefont {Beyer}\ \emph {et~al.}(2014)\citenamefont {Beyer},
  \citenamefont {Daszuta}, \citenamefont {Frauendiener},\ and\ \citenamefont
  {Whale}}]{Beyer:2013loa}%
  \BibitemOpen
  \bibfield  {author} {\bibinfo {author} {\bibfnamefont {F.}~\bibnamefont
  {Beyer}}, \bibinfo {author} {\bibfnamefont {B.}~\bibnamefont {Daszuta}},
  \bibinfo {author} {\bibfnamefont {J.}~\bibnamefont {Frauendiener}}, \ and\
  \bibinfo {author} {\bibfnamefont {B.}~\bibnamefont {Whale}},\ }\bibfield
  {title} {\enquote {\bibinfo {title} {{Numerical evolutions of fields on the
  2-sphere using a spectral method based on spin-weighted spherical
  harmonics}},}\ }\href {\doibase 10.1088/0264-9381/31/7/075019} {\bibfield
  {journal} {\bibinfo  {journal} {Classical Quantum Gravity}\ }\textbf
  {\bibinfo {volume} {31}},\ \bibinfo {pages} {075019} (\bibinfo {year}
  {2014})},\ \Eprint {http://arxiv.org/abs/1308.4729} {arXiv:1308.4729
  [physics.comp-ph]} \BibitemShut {NoStop}%
\bibitem [{\citenamefont {Ruiz}\ \emph {et~al.}(2008)\citenamefont {Ruiz},
  \citenamefont {Takahashi}, \citenamefont {Alcubierre},\ and\ \citenamefont
  {Nunez}}]{Ruiz:2007yx}%
  \BibitemOpen
  \bibfield  {author} {\bibinfo {author} {\bibfnamefont {M.}~\bibnamefont
  {Ruiz}}, \bibinfo {author} {\bibfnamefont {R.}~\bibnamefont {Takahashi}},
  \bibinfo {author} {\bibfnamefont {M.}~\bibnamefont {Alcubierre}}, \ and\
  \bibinfo {author} {\bibfnamefont {D.}~\bibnamefont {Nunez}},\ }\bibfield
  {title} {\enquote {\bibinfo {title} {{Multipole expansions for energy and
  momenta carried by gravitational waves}},}\ }\href {\doibase
  10.1007/s10714-007-0570-8, 10.1007/s10714-008-0684-7} {\bibfield  {journal}
  {\bibinfo  {journal} {Gen. Relativ. Gravit.}\ }\textbf {\bibinfo {volume}
  {40}},\ \bibinfo {pages} {2467} (\bibinfo {year} {2008})},\ \Eprint
  {http://arxiv.org/abs/0707.4654} {arXiv:0707.4654 [gr-qc]} \BibitemShut
  {NoStop}%
\bibitem [{\citenamefont {Campiglia}\ and\ \citenamefont
  {Laddha}(2014)}]{Campiglia:2014yka}%
  \BibitemOpen
  \bibfield  {author} {\bibinfo {author} {\bibfnamefont {M.}~\bibnamefont
  {Campiglia}}\ and\ \bibinfo {author} {\bibfnamefont {A.}~\bibnamefont
  {Laddha}},\ }\bibfield  {title} {\enquote {\bibinfo {title} {{Asymptotic
  symmetries and subleading soft graviton theorem}},}\ }\href {\doibase
  10.1103/PhysRevD.90.124028} {\bibfield  {journal} {\bibinfo  {journal} {Phys.
  Rev. D}\ }\textbf {\bibinfo {volume} {90}},\ \bibinfo {pages} {124028}
  (\bibinfo {year} {2014})},\ \Eprint {http://arxiv.org/abs/1408.2228}
  {arXiv:1408.2228 [hep-th]} \BibitemShut {NoStop}%
\bibitem [{\citenamefont {Campiglia}\ and\ \citenamefont
  {Laddha}(2015)}]{Campiglia:2015yka}%
  \BibitemOpen
  \bibfield  {author} {\bibinfo {author} {\bibfnamefont {M.}~\bibnamefont
  {Campiglia}}\ and\ \bibinfo {author} {\bibfnamefont {A.}~\bibnamefont
  {Laddha}},\ }\bibfield  {title} {\enquote {\bibinfo {title} {{New symmetries
  for the Gravitational S-matrix}},}\ }\href {\doibase 10.1007/JHEP04(2015)076}
  {\bibfield  {journal} {\bibinfo  {journal} {J. High Energy Phys.}\ }\textbf
  {\bibinfo {volume} {04}},\ \bibinfo {pages} {076} (\bibinfo {year} {2015})},\
  \Eprint {http://arxiv.org/abs/1502.02318} {arXiv:1502.02318 [hep-th]}
  \BibitemShut {NoStop}%
\bibitem [{\citenamefont {Blanchet}(1997)}]{Blanchet:1996vx}%
  \BibitemOpen
  \bibfield  {author} {\bibinfo {author} {\bibfnamefont {L.}~\bibnamefont
  {Blanchet}},\ }\bibfield  {title} {\enquote {\bibinfo {title} {{Gravitational
  radiation reaction and balance equations to post-Newtonian order}},}\ }\href
  {\doibase 10.1103/PhysRevD.55.714} {\bibfield  {journal} {\bibinfo  {journal}
  {Phys. Rev. D}\ }\textbf {\bibinfo {volume} {55}},\ \bibinfo {pages}
  {714--732} (\bibinfo {year} {1997})},\ \Eprint
  {http://arxiv.org/abs/gr-qc/9609049} {arXiv:gr-qc/9609049 [gr-qc]}
  \BibitemShut {NoStop}%
\bibitem [{\citenamefont {Ajith}(2011)}]{Ajith:2011ec}%
  \BibitemOpen
  \bibfield  {author} {\bibinfo {author} {\bibfnamefont {P.}~\bibnamefont
  {Ajith}},\ }\bibfield  {title} {\enquote {\bibinfo {title} {{Addressing the
  spin question in gravitational-wave searches: Waveform templates for
  inspiralling compact binaries with nonprecessing spins}},}\ }\href {\doibase
  10.1103/PhysRevD.84.084037} {\bibfield  {journal} {\bibinfo  {journal} {Phys.
  Rev. D}\ }\textbf {\bibinfo {volume} {84}},\ \bibinfo {pages} {084037}
  (\bibinfo {year} {2011})},\ \Eprint {http://arxiv.org/abs/1107.1267}
  {arXiv:1107.1267 [gr-qc]} \BibitemShut {NoStop}%
\bibitem [{\citenamefont {Regimbau}\ \emph {et~al.}(2012)\citenamefont
  {Regimbau} \emph {et~al.}}]{Regimbau:2012ir}%
  \BibitemOpen
  \bibfield  {author} {\bibinfo {author} {\bibfnamefont {T.}~\bibnamefont
  {Regimbau}} \emph {et~al.},\ }\bibfield  {title} {\enquote {\bibinfo {title}
  {{A Mock Data Challenge for the Einstein Gravitational-Wave Telescope}},}\
  }\href {\doibase 10.1103/PhysRevD.86.122001} {\bibfield  {journal} {\bibinfo
  {journal} {Phys. Rev. D}\ }\textbf {\bibinfo {volume} {86}},\ \bibinfo
  {pages} {122001} (\bibinfo {year} {2012})},\ \Eprint
  {http://arxiv.org/abs/1201.3563} {arXiv:1201.3563 [gr-qc]} \BibitemShut
  {NoStop}%
\bibitem [{\citenamefont {Abbott}\ \emph
  {et~al.}(2016{\natexlab{b}})\citenamefont {Abbott} \emph
  {et~al.}}]{Abbott:2016blz}%
  \BibitemOpen
  \bibfield  {author} {\bibinfo {author} {\bibfnamefont {B.~P.}\ \bibnamefont
  {Abbott}} \emph {et~al.} (\bibinfo {collaboration} {Virgo, LIGO
  Scientific}),\ }\bibfield  {title} {\enquote {\bibinfo {title} {{Observation
  of Gravitational Waves from a Binary Black Hole Merger}},}\ }\href {\doibase
  10.1103/PhysRevLett.116.061102} {\bibfield  {journal} {\bibinfo  {journal}
  {Phys. Rev. Lett.}\ }\textbf {\bibinfo {volume} {116}},\ \bibinfo {pages}
  {061102} (\bibinfo {year} {2016}{\natexlab{b}})},\ \Eprint
  {http://arxiv.org/abs/1602.03837} {arXiv:1602.03837 [gr-qc]} \BibitemShut
  {NoStop}%
\bibitem [{\citenamefont {Santamaria}\ \emph {et~al.}(2010)\citenamefont
  {Santamaria} \emph {et~al.}}]{Santamaria:2010yb}%
  \BibitemOpen
  \bibfield  {author} {\bibinfo {author} {\bibfnamefont {L.}~\bibnamefont
  {Santamaria}} \emph {et~al.},\ }\bibfield  {title} {\enquote {\bibinfo
  {title} {{Matching post-Newtonian and numerical relativity waveforms:
  systematic errors and a new phenomenological model for non-precessing black
  hole binaries}},}\ }\href {\doibase 10.1103/PhysRevD.82.064016} {\bibfield
  {journal} {\bibinfo  {journal} {Phys. Rev. D}\ }\textbf {\bibinfo {volume}
  {82}},\ \bibinfo {pages} {064016} (\bibinfo {year} {2010})},\ \Eprint
  {http://arxiv.org/abs/1005.3306} {arXiv:1005.3306 [gr-qc]} \BibitemShut
  {NoStop}%
\bibitem [{\citenamefont {Lommen}(2015)}]{Lommen:2015gbz}%
  \BibitemOpen
  \bibfield  {author} {\bibinfo {author} {\bibfnamefont {A.~N.}\ \bibnamefont
  {Lommen}},\ }\bibfield  {title} {\enquote {\bibinfo {title} {{Pulsar timing
  arrays: the promise of gravitational wave detection}},}\ }\href {\doibase
  10.1088/0034-4885/78/12/124901} {\bibfield  {journal} {\bibinfo  {journal}
  {Rep. Prog. Phys.}\ }\textbf {\bibinfo {volume} {78}},\ \bibinfo {pages}
  {124901} (\bibinfo {year} {2015})}\BibitemShut {NoStop}%
\end{thebibliography}%

\end{document}